 \documentclass[aps, twocolumn, showpacs, amsmath]{revtex4}
\usepackage{dcolumn}
\usepackage{bm}
\usepackage{graphicx}
\usepackage{color}
\usepackage{subfigure}
\usepackage{hyperref}
\usepackage{latexsym}
\usepackage{amsthm}
\usepackage{amssymb}
\usepackage{braket}
\DeclareGraphicsExtensions{.jpg,.pdf, .mps, .png, .eps, .ps, .EPS,.gif}

\begin{document}
\def\be{\begin{equation}}
\def\ee{\end{equation}}

\def\bc{\begin{center}}
\def\ec{\end{center}}
\def\bea{\begin{eqnarray}}
\def\eea{\end{eqnarray}}
\newcommand{\avg}[1]{\langle{#1}\rangle}
\newcommand{\Avg}[1]{\left\langle{#1}\right\rangle}

\def\ie{\textit{i.e.}}
\def\etal{\textit{et al.}}
\def\m{\vec{m}}
\def\G{\mathcal{G}}

\newcommand{\davide}[1]{{\bf\color{blue}#1}}
\newcommand{\gin}[1]{{\bf\color{green}#1}}

\title{ Explosive higher-order Kuramoto dynamics on simplicial complexes}
\author{Ana P. Mill\'an} 
\address{
Department of Clinical Neurophysiology and MEG Center, 
Amsterdam UMC, Vrije Universiteit Amsterdam,  Amsterdam,  1081 HV, The Netherlands}
\author{Joaqu\'in J. Torres}
\address{Departamento de Electromagnetismo y F\'isica  de la Materia
and Instituto Carlos I de F\'isica Te\'orica y Computacional, Universidad de Granada, 18071 Granada, Spain}
\author{Ginestra Bianconi}  

\address{School of Mathematical Sciences, Queen Mary University of London, E1 4NS London, United Kingdom\\
Alan Turing Institute, The British Library, London, United Kingdom}
\begin{abstract}
The higher-order interactions of complex systems, such as the brain  are captured by their simplicial complex structure and have a significant effect on dynamics. 
However, the existing dynamical models defined on simplicial complexes make the strong assumption that the dynamics resides exclusively on the nodes. Here we formulate the higher-order Kuramoto model which describes the interactions between oscillators placed not only on nodes but also on links, triangles, and so on. 
We show that higher-order Kuramoto dynamics can lead to an explosive synchronization transition by using an adaptive coupling dependent on the solenoidal and the irrotational component of the dynamics.
\end{abstract}
\maketitle

From the brain \cite{Bassett,Petri,BlueBrain,benson,santos} to   social interactions \cite{Dane,Latora,Kahng,Arenas_epidemics} and complex materials \cite{Bassett_granular,Tadic}, a vast number of complex systems  have the underlying topology of  simplicial complexes \cite{Perspective,Lambiotte,Porter}. 
Simplicial complexes are topological structures  formed by simplices of different dimension such as nodes, links, triangles, tetrahedra and so on, and capture the many-body interactions between the elements of an interacting complex system.
In the last years simplicial complex modelling has attracted  significant attention  \cite{Farber,Configuration,Hyperbolic,Petri_dynamical} revealing  the fundamental mechanisms determining emergent network geometry  \cite{Emergent} and the interplay between network geometry and degree correlations \cite{Configuration}.
Modelling complex systems using simplicial complexes 
allows for the very fertile perspective of considering the role that higher-order interactions have on dynamical processes.  
For instance, recent works  \cite{Ziff,Dane,Latora,Kahng,Arenas_epidemics,Ana,Ana2,Skardal1,Skardal2} on simplicial complex dynamics, including works on simplicial complex synchronization \cite{Ana,Ana2,Skardal1,Skardal2}, reveal  that the  topology and  geometry of the simplicial complexes and their many-body interactions induce cooperative phenomena that cannot be found in pairwise networks.

In the last  years explosive  synchronization \cite{explosive1,explosive2} is  attracting increasing scientific interest.  Different pathways to explosive synchronization have been explored in the framework of the Kuramoto dynamics of single and multilayer networks. 
These notably include correlating the intrinsic frequency of the nodes to their degree \cite{Arenasdegree} or modulating the  coupling between different oscillators adaptively using the local order parameter  in single networks and in multiplex networks \cite{Boccaletti,NaturePhysics}. 
An outstanding  open question is to establish  the conditions that allow explosive synchronization on simplicial complexes.

Among the papers investigating  synchonization dynamics beyond pairwise interactions \cite{Tanaka,Pikovsky},  recent works \cite{Bick,Skardal2} have  proposed a many-body Kuramoto model where the phases associated with the nodes of the network can be coupled in triplets or quadruplets if the corresponding nodes share a triangle or a tetrahedron. Interestingly in this context it has been shown \cite{Skardal2} that the many-body Kuramoto dynamics can lead to explosive, i.e. discontinuous phase transitions. 
However,  the vast majority of works that address the study of dynamics on simplicial complexes,   have the  limitation that they  associate a dynamic variable exclusively with nodes of a network. 
Here we are interested in a much more general scenario where the dynamics can be associated with the faces of  dimension $n\geq 0$ of a simplicial complex.  
Indeed dynamical processes might not just reside on nodes, instead they might be  related directly to dynamics defined  on higher dimensional simplices  leading  to the definition of   topological dynamical signals \cite{Barbarossa}. 
For instance  each link  can be associated with a flux. 
Flow dynamics is relevant  for biological transport networks including  fungal networks \cite{Tero}, tree vascular networks \cite{Katifori}, microvascular networks \cite{alim}, or hemodynamic in the mammalian cortex \cite{cortex},  where there is  some evidence that  the dynamics can spontaneously give rise to  oscillatory currents. Flow signals can be also used to analyze fMRI brain data \cite{DeVille} and to study blood flow between different regions of the brain.
More in general,  the simplicial complex can be considered as a representation of interactions of different order. For instance,   for any given networked structure the line-graph construction \cite{linegraph1,linegraph2} allows to map links into nodes of the line-graph, so that a dynamics defined on the links of a simplicial complex can be mapped on a node dynamics of its line graph. However,  the original simplicial complex provides a definition of the  many-body interactions solidly based on  topology.

In this Letter we formulate a higher-order Kuramoto dynamics where the dynamical variables are coupled oscillators associated with higher dimensional simplices such as nodes, links, triangles and so on.
By using Hodge decomposition  we show that the dynamics defined on an $n$-dimensional simplex can be projected on the dynamics defined on $n+1$ and $n-1$ dimensional simplices. We propose a simple higher-order Kuramoto dynamics in which these two projected dynamics are decoupled and display a  continuous phase transition.
We then formulate the explosive higher-order Kuramoto dynamics which adaptively couples the two projected dynamics with a mechanism inspired by Ref. \cite{Boccaletti}, showing that in this case the explosive higher-order Kuramoto dynamics leads to a discontinuous synchronization transition.
 This implies for instance that a dynamics defined on links can induce a simultaneous explosive synchronization on the dynamics projected  on nodes and triangles.
Therefore our work elucidates an important mechanism leading to higher-order explosive Kuramoto dynamics. 

\begin{figure}[h!]
 \includegraphics[width=0.5\columnwidth]{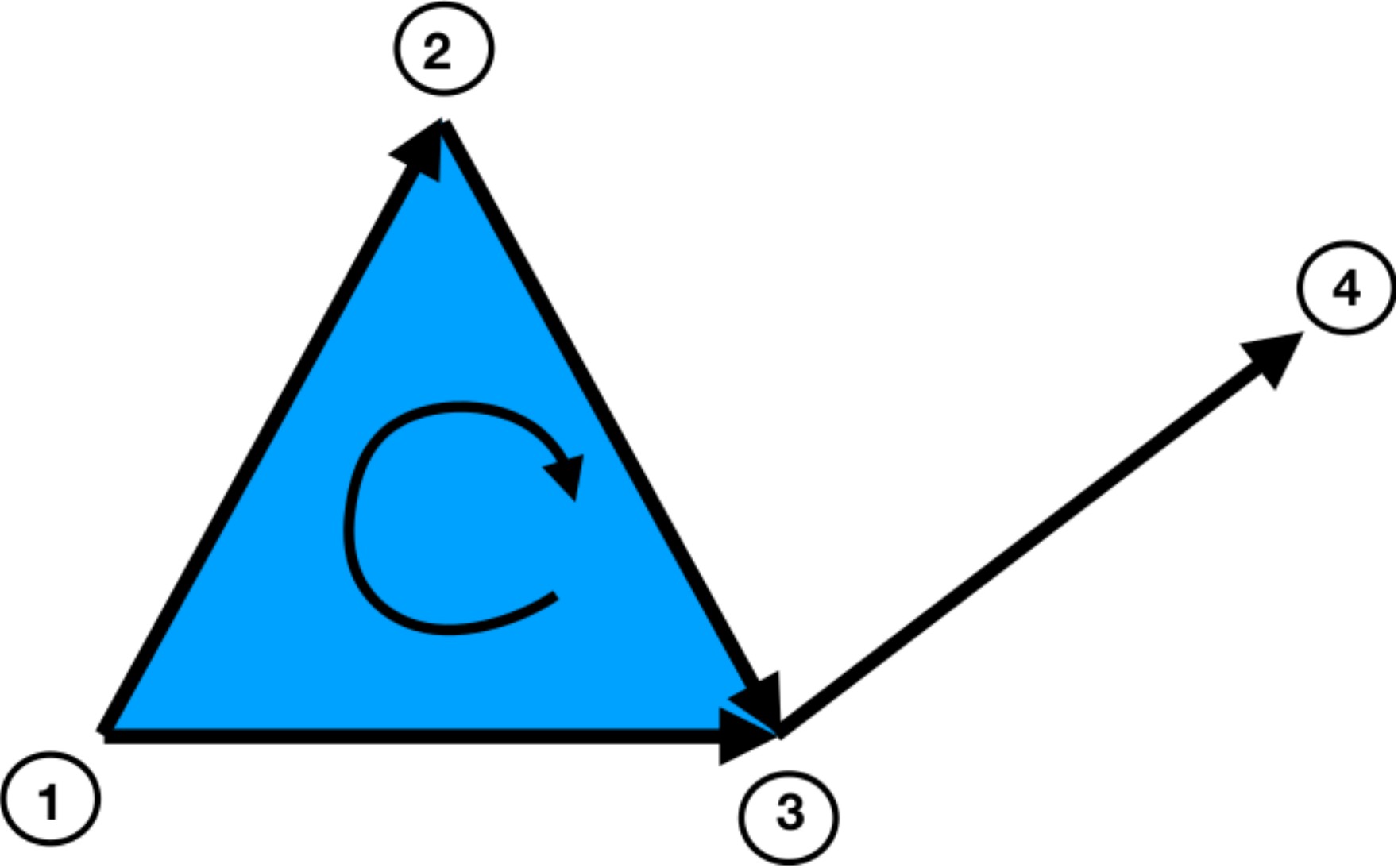}	\caption{ An example of a small simplicial complex with the orientation of the simplices induced by the labelling of the nodes.}
	\label{fig:example}
	\end{figure}

 {\it Definition of simplicial complexes -}  Simplicial complexes represent higher-order networks, which include interactions between two or more nodes, described by simplices. 
 A node is a $0$-dimensional simplex, a link is a $1$-dimensional simplex, a triangle is a $2$-dimensional simplex, a tetrahedron is a $3$-dimensional simplex, and so on. 
 The faces of a  simplex $\alpha$ of dimension $n$ are all the simplices $\alpha^{\prime}$ of dimension $n^{\prime}<n$ that can be constructed by taking proper subsets of the set of all the nodes forming the simplex $\alpha$. 
 A simplicial complex ${\mathcal K}$ is  formed by a set of simplices that satisfy the condition of closure (given a simplex belonging to the simplicial complex all its faces also belong to the simplicial complex). 
 In this work we will use the configuration model \cite{Configuration} of simplicial complexes, which naturally generalizes the configuration model of networks (see  Supplemental Material (SM) for other topologies). 
 In particular, the $d$-dimensional configuration model generates simplicial complexes formed by gluing $d$-dimensional simplices such that every node is incident to a given number of $d$-dimensional simplices called its generalized degree.
 In topology simplices have also an orientation. A $n$-dimensional {\it oriented simplex} $\alpha$ is a set of ordered  $n+1$ nodes
$$\alpha=[i_0,i_1,\dots,i_{n}].$$ 
For instance, a link $\alpha=[i,j]$ has opposite sign of the link $[j,i]$, i.e.
$[i,j]=-[j,i].$
Similarly, we associate an orientation to higher-order simplices satisfying
\bea
[i_0,i_1,\ldots, i_n]=(-1)^{\sigma(\pi)}[i_{\pi(0)},i_{\pi(1)},\dots,i_{\pi(n)}],
\eea
where $\sigma(\pi)$ indicates the parity of the permutation ${\bf \pi}$.
Here we consider the orientation induced by the labelling of its nodes, i.e. for every simplex in a simplicial complex we give positive orientation as the one provided by the increasing list of node labels (see Figure $\ref{fig:example}$).

In topology \cite{Goldberg,Egerstedt,Jost,Barbarossa}, the $n$-chains ${\mathcal C}_n$ are the elements of a free abelian group with basis the $n$-dimensional simplices of a simplicial complex. The boundary map is a linear map $\partial_n:{\mathcal C}_n\to {\mathcal C}_{n-1}$ defined by its action on each simplex. 
Specifically, the boundary map  maps every  $n$-dimensional simplex $\alpha$ to a linear combination of the $(n-1)$-dimensional oriented faces at its boundary, given by 
\bea
\partial_n [i_0,i_1\ldots,i_n]=\sum_{p=0}^n(-1)^p[i_0,i_1,\dots,i_{p-1},i_{p+1},\dots,i_n].
\eea
The boundary map satisfies the important property that 
$\partial_{n-1}\partial_n=0$,
that is usually expressed by saying that the boundary of a boundary is null (see SM \cite{SI}).
Given a simplicial complex with $N_{[n]}$ $n$-dimensional simplices,  the boundary map $\partial_n$ can be described using the $N_{[n-1]}\times N_{[n]}$ incidence matrix ${\bf B}_{[n]}$ (see SM \cite{SI}).
For instance, in Figure $\ref{fig:example}$ we show an example  of a simplicial complex formed by the set of nodes $\{[1],[2],[3],[4]\}$, the set of links $\{[1,2],[1,3],[2,3],[3,4]\}$, and the set of triangles $\{[1,2,3]\}$. 
The incidence matrices  \cite{Goldberg,Egerstedt,Jost,Barbarossa} of this simplicial complex are given by 
\bea
{\bf B}_{[1]}=\begin{array}{c|cccc}
&$[1,2]$&$[1,3]$&$[2,3]$&$[3,4]$\\
\hline
$[1]$&-1&-1 &0&0\\
$[2]$&1&0&-1&0\\
$[3]$&0&1&1&-1\\
$[4]$&0&0&0&1\\
\end{array},
\ 
{\bf B}_{[2]}=\begin{array}{c|c}
 &$[1,2,3]$\\
 \hline
$[1,2]$&1\\
$[1,3]$&-1\\
$[2,3]$&1\\
$[3,4]$&0
\end{array}.
\eea

{\it Higher-order Laplacians -} The graph Laplacian is widely used to study dynamical processes defined on the nodes of a network. It can be expressed in terms of the boundary matrix ${\bf B}_{[1]}$ as
\bea
{\bf L}_{[0]}={\bf B}_{[1]}{\bf B}^{\top}_{[1]}.
\eea
The higher-order Laplacian  ${\bf L}_{[n]}$ \cite{Goldberg,Egerstedt,Jost,Barbarossa}, with $n>0$,   generalizes the  graph Laplacian by describing diffusion taking place on $n$ dimensional faces. 
The $n$-th Laplacian ${\bf L}_{[n]}$ is  an  $N_{[n]}\times N_{[n]}$ matrix given by
\bea
{\bf L}_{[n]}={\bf B}^{\top}_{[n]}{\bf B}_{[n]}+{\bf B}_{[n+1]}{\bf B}^{\top}_{[n+1]}.
\eea
The spectral properties of the higher-order Laplacian can be proven to be independent of the orientation of the simplices as long as the orientation is induced by a labelling of the nodes.
The main property of the higher-order Laplacian 
 is that the degeneracy of the zero eigenvalue of 
 ${\bf L}_{[n]}$ is equal to the Betti number $\beta_n$, and that its corresponding eigenvectors localize around the corresponding $n$-dimensional cavities of the simplicial complex.

The higher-order Laplacians have notable spectral properties induced by the topological properties of the boundary map \cite{Barbarossa}. 
In fact, given that $\partial_{n-1}\partial_n=0$, we have ${\bf B}_{[n-1]}{\bf B}_{[n]}={\bf 0}$ and, similarly,  ${\bf B}^{\top}_{[n]}{\bf B}^{\top}_{[n-1]}={\bf 0}$.
Therefore the eigenvectors associated with the non-null eigenvalues of ${\bf L}_{[n]}^{[up]}={\bf B}_{[n+1]}{\bf B}^{\top}_{[n+1]}$ are orthogonal to the eigenvectors associated with the non-null eigenvalues of 
${\bf L}_{[n]}^{[down]}={\bf B}^{\top}_{[n]}{\bf B}_{[n]}$. 
It follows that  the non-null eigenvalues of ${\bf L}_{[n]}$ are either the non-null eigenvalues of ${\bf L}^{[up]}_{[n]}$ or the non-null eigenvalues of  ${\bf L}^{[down]}_{[n]}$.
This property of the higher-order Laplacian can be exploited to prove that every vector ${\bf x}_{[n]}$ defined on $n$-dimensional simplices can be decomposed according to the {\em Hodge decomposition} \cite{Barbarossa} as 
\bea
{\bf x}_{[n]}={\bf x}_{[n]}^{H}+{\bf B}_{[n]}^{\top}{\bf z}_{[n-1]}+{\bf B}_{[n+1]}{\bf z}_{[n+1]}
\eea
where ${\bf x}_{[n]}^{H}$ is the harmonic component that satisfies ${\bf B}_{[n+1]}^{\top}{\bf x}_{[n]}^{H}=0$, ${\bf B}_{[n]}{\bf x}_{[n]}^{H}={\bf 0}$,  
the term ${\bf B}_{[n]}^{\top}{\bf z}_{[n-1]}$ is the irrotational component as we have ${\bf B}_{[n+1]}^{\top}{\bf B}_{[n]}^{\top}{\bf z}_{[n-1]}={\bf 0}$ and the third term ${\bf B}_{[n+1]}{\bf z}_{[n+1]}$ is the solenoidal component as we have ${\bf B}_{[n]}{\bf B}_{[n+1]}{\bf z}_{[n+1]}={\bf 0}$.

{\it Higher-order Kuramoto dynamics -}   
The Kuramoto model  \cite{Kuramoto} is a dynamical model for the vector  $\bm{\theta}$ whose elements are the phases $\theta_i$ associated to the nodes of the simplicial complex. Each oscillator $i$ has   an internal frequency $\omega_i$ and is coupled pairwise to the oscillator of the connected nodes by the coupling constant $\sigma$. 
Interestingly, the Kuramoto dynamics can be interpreted as a dynamics defined on the nodes of a simplicial complex, i.e. simplices of dimension $n=0$, indicated with label $i=1,2,\ldots, N_{[0]}$, and it can be expressed in terms of the incidence matrix ${\bf B}_{[1]}$ (see SI) as
\bea
\dot {\bm{\theta}}&=&\bm{\omega}-\sigma  {\bf B}_{[1]}\sin {\bf B}^{\top}_{[1]}\bm{\theta},
\eea
 where here and in the following   $\sin {\bf x}$ indicates the column vector where the sine function is taken element wise and $\bm{\omega}$ is the vector of internal frequencies $\omega_i$ associated to the nodes of the simplicial complex.

Here our goal is to extend the Kuramoto dynamics to describe synchronization among dynamical phases $\theta_{\alpha}$ associated to each simplex $\alpha$ of dimension $n>0$, i.e.  links (for $n=1)$ or even higher dimensional simplices.  We assume that these dynamical signals are phases that  oscillate with some internal frequency and they can be coupled by higher-order interactions.  The natural way to choose the coupling between  $n$-dimensional phases is suggested by the generalization of the Kuramoto dynamics using the   higher-order incidence matrices 
\bea
\dot {\bm{\theta}}&=&\bm{\omega}-\sigma  {\bf B}_{[n+1]}\sin {\bf B}^{\top}_{[n+1]}\bm{\theta}-\sigma  {\bf B}^{\top}_{[n]}\sin {\bf B}_{[n]}\bm{\theta},
\eea
where $\bm{\theta}$ is the vector of phases $\theta_{\alpha}$ and  where $\bm{\omega}$ is the vector of  intrinsic frequencies $\omega_{\alpha}$  associated to each $n$-dimensional simplex $\alpha$.  Each internal frequency $\omega_{\alpha}$ is  drawn from a normal distribution with mean $\Omega$ and variance $1$, i.e. $\omega\sim{\mathcal N}(\Omega,1)$. 
The higher-order Kuramoto dynamics describes a dynamics of phases associated to simplices of dimension $n$ as links ($n=1$), triangles ($n=2$) and so on (see SM \cite{SI}).

An important question to ask is whether the dynamics associated to $n$-dimensional simplices induces a dynamics on  lower or higher dimensional  simplices. For instance, if we have a Kuramoto dynamics defined on links, what is  the effect of this dynamics on nodes and triangles?
It turns out that there is a simple way to project the dynamics defined on links into dynamics defined on nodes and triangles suggested by topology. More in general, we can project the dynamics defined on $n$ simplices to the dynamics defined on simplices of dimension $n-1$ and $n+1$ by using the higher-order incidence matrices.
To this end, let us indicate with  $\bm{\theta}^{[+]}$ the  vector of $N_{[n+1]}$ phases associated to each $n+1$ simplex of the simplicial complex. 
This vector   describes the projection of the dynamics on simplices of dimension  $n+1$.
Similarly, let us indicate with $\bm{\theta}^{[-]}$  the vector of $N_{[n-1]}$ phases associated with each $n-1$ simplex of the simplicial complex. 
This vector represents the projection of the dynamics on simplices of dimension  $n-1$.
Topological considerations  suggest  the physical meaning of these projecting as $\bm{\theta}^{[+]}$ and $\bm{\theta}^{[-]}$ are respectively as the ``discrete curl" and ``discrete divergence" of $\bm{\theta}$ i.e. 
\bea
{\bm{\theta}^{[+]}}&=&{\bf B}^{\top}_{[n+1]}\bm{\theta},\nonumber \\
{\bm{\theta}}^{[-]}&=&{\bf B}_{[n]}\bm{\theta}.
\eea
Using the Hodge decomposition it is easy to  show that ${\bm{\theta}^{[+]}}$  depends only  on the solenoidal component of the dynamics defined on $n$-dimensional phases, whereas ${\bm{\theta}^{[-]}}$  depends only on the irrotational  component. 
Since we have that ${\bf B}^{\top}_{[n]}{\bf B}^{\top}_{[n-1]}={\bf 0}$ and ${\bf B}_{[n-1]}{\bf B}_{[n]}={\bf 0}$,  if $\bm{\theta}$ obeys the higher-Kuramoto dynamics,  then the projected dynamical variables ${\bm{\theta}^{[+]}}$  and ${\bm{\theta}^{[-]}}$ evolve  independently according to 
\bea
{{\dot{\bm{\theta}}^{[+]}}}&=&{\bf B}^{\top}_{[n+1]}\bm{\omega}-\sigma{\bf L}_{[n+1]}^{[down]}\sin ({\bm{\theta}}^{[+]}),\nonumber \\
{{\dot{\bm{\theta}}^{[-]}}}&=&{\bf B}_{[n]}\bm{\omega}-\sigma{\bf L}_{[n-1]}^{[up]}\sin ({\bm{\theta}}^{[-]}).
\eea
Therefore, the dynamics defined on $n$-dimensional simplices can naturally be decoupled into two non-interacting dynamics acting on $n-1$ and on $n+1$ dimensional simplices.
The order parameter for each of these two independent dynamics are
$R^{[+]}=\left|\sum_{\alpha=1}^{N_{[n+1]}}e^{i\theta^{[+]}_\alpha}\right|/{N_{[n+1]}}$,
$R^{[-]}=\left|\sum_{\alpha=1}^{N_{[n-1]}}e^{i\theta^{[-]}_\alpha}\right|/N_{[n-1]}.$

In order to investigate the properties of the dynamics defined on $n$-dimensional simplices, we can consider the standard order parameter $R$ given by 
$R=\left|\sum_{\alpha=1}^{N_{[n]}}e^{i \theta_{\alpha}}\right|/N_{[n]}$ 
and two additional order parameters
$R^{[1]}=\left|\sum_{\alpha=1}^{N_{[n]}}e^{i y_{\alpha}^{[1]}}\right|/N_{[n]},$ and $R^{[2]}=\left|\sum_{\alpha=1}^{N_{[n]}}e^{i y_{\alpha}^{[2]}}\right|/N_{[n]},$
where ${\bf{y}^{[1]}}={\bf L}_{[n]}^{[up]}\bm{\theta}={\bf L}_{[n]}^{[up]}{\bf B}_{[n+1]}{\bf z}_{[n+1]}$ depends only on the solenoidal component of the dynamics on $n$-dimensional simplices, and
${\bf{y}}^{[2]}={\bf L}_{[n]}^{[down]}\bm{\theta}={\bf L}_{[n]}^{[down]}{\bf B}_{[n]}^{\top}{\bf z}_{[n-1]}$ depends only on the irrotational component of the dynamics on $n$-dimensional simplices.

We have simulated the higher-order {$(n=1)$} Kuramoto dynamics on the $3$-dimensional simplicial complexes produced by the configuration model with power-law generalized degree distribution of the nodes.
These simplicial complexes have Betti numbers $\beta_1>0,\beta_2=0$.
 We observe that the projected dynamics on the $2$-dimensional simplices and the $0$-dimensional simplices display a continuous  synchronization transition (see  Fig. $\ref{fig:simple}$).
When we investigate the three order parameters for the dynamics defined on $n$-dimensional simplices we observe that $R$ does not capture the collective behavior of the phases due to the fact that the harmonic component of their dynamics is not coupled by the higher-order Kuramoto dynamics. However the order parameters $R^{[1]}$ and $R^{[2]}$ are sensible to the synchronization of the solenoidal and irrotational component of the dynamics of the phases (see blue points in Fig. $\ref{fig:combined}$).  This suggests that this ordering in physical system can go un-noticed if the correct order parameters are not applied to the signal. A phenomenological analytical approach can show that while the projection of the phases on the harmonic modes are decoupled $\bm{\theta}^{[+]}$ and $\bm{\theta}^{[-]}$ have a continuous synchronization transition at $\sigma_c=0$ (see SM \cite{SI}).
The nature of the phase transition does not  change if we consider simplicial complexes with Poisson generalized degree distribution of the nodes and can be explained by an analytical framework (see SM \cite{SI}) 

	\begin{figure}[h!]
 \includegraphics[width=1\columnwidth]{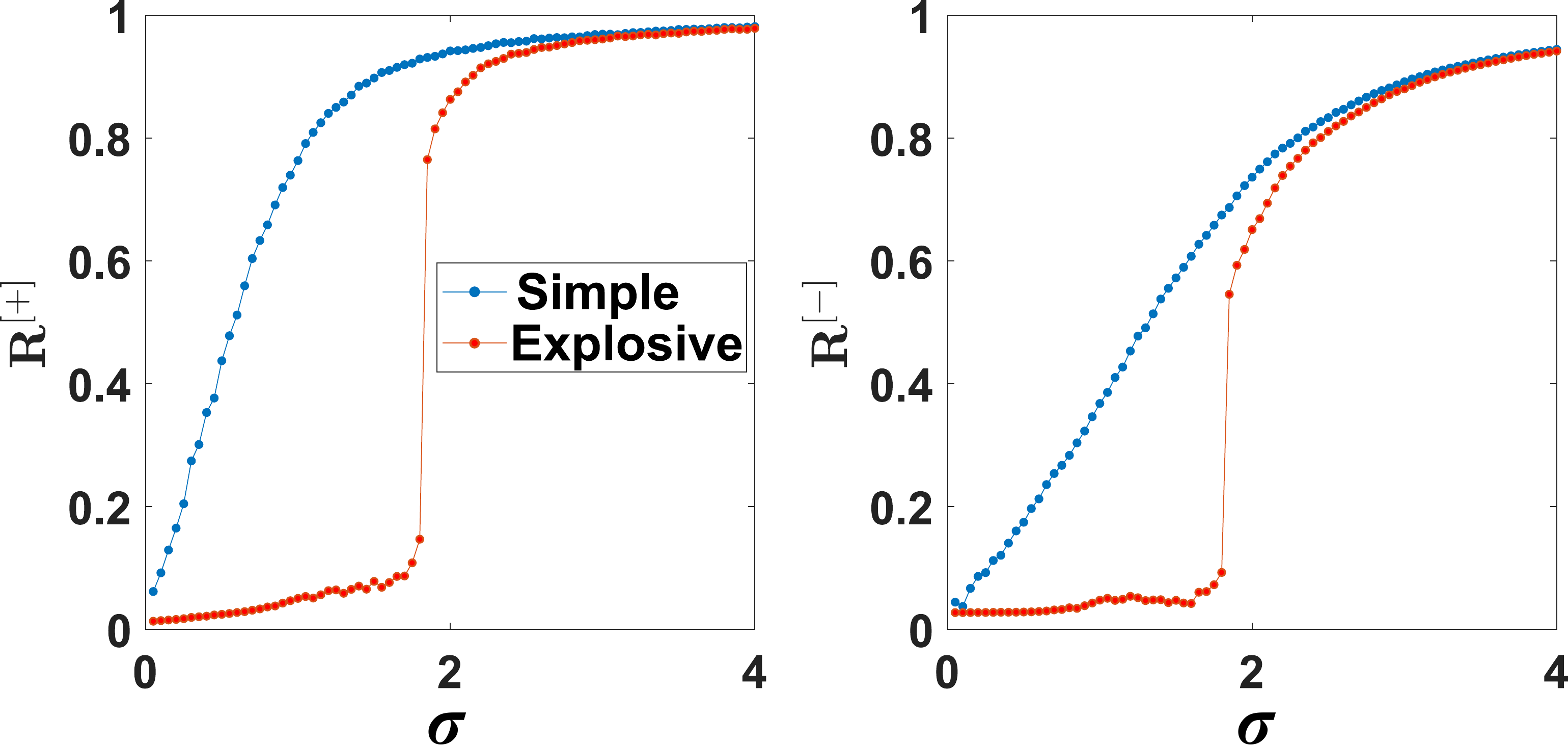}	
 \caption{The projection of the  higher-order ($n=1$) Kuramoto dynamics on $(n-1)$-dimensional faces and $(n+1)$-dimensional faces is investigated by plotting the order parameters $R^{[+]}$ (left panel) and $R^{[-]}$ (right panel), both for the simple (blue circles) and explosive (red squares) dynamics.
 Here  both the simple and the explosive higher-order Kuramoto model have   $\Omega=2$ and are defined on a configuration model of $N_{[0]}=1000$ nodes, $N_{[1]}=5299$ links and $N_{[2]}=4147$ triangles with generalized degree of the nodes that is power-law distributed with power-law exponent $\gamma=2.8$. 
  }
	\label{fig:simple}\label{fig:coupled}
	\end{figure}

		\begin{figure}[h!]
	 \includegraphics[width=1\columnwidth]{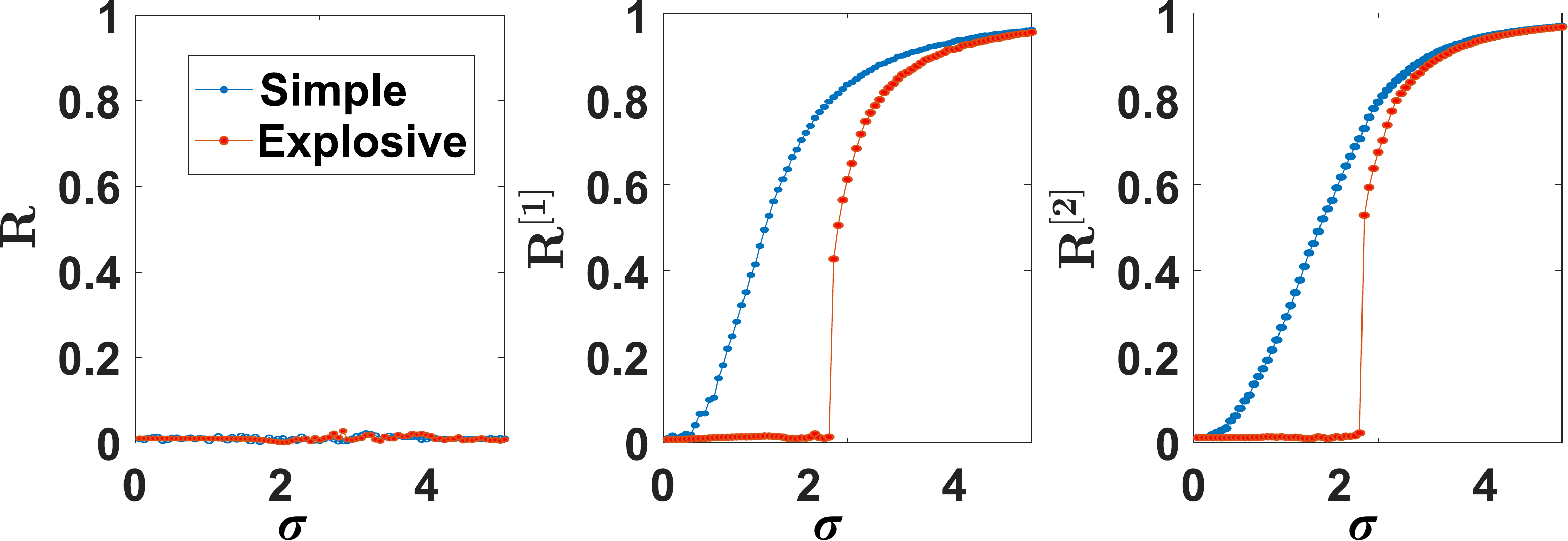}
	 \caption{The order parameters $R$, $R^{[1]}$ and $R^{[2]}$ of  the simple (blue circles) and explosive (red squares) higher-order ($n=1$)  Kuramoto dynamics  are plotted versus the coupling constant $\sigma$.  The network parameters are the same as in Fig. $\ref{fig:simple}$. }
	\label{fig:combined}
	\end{figure}

{\it Explosive higher-order Kuramoto dynamics -}
In order  to explore whether it is possible to enforce an explosive phase transition we include a coupling between the equations determining the dynamics of ${\bm{\theta}^{[+]}}$ and ${\bm{\theta}^{[-]}}$. 
The way we coupled these two independent dynamics is inspired by the coupling of the dynamics of multiplex Kuramoto dynamics in Ref. \cite{Boccaletti}.
However, while in the explosive multiplex Kuramoto dynamics the coupling between the phases in one layer is modulated by the  local order parameter of each node in the other layer, here we consider a modulation of the coupling between the phases $\bm{\theta^{[+]}}$ and $\bm{\theta^{[-]}}$ given respectively by the global order parameters $R^{[-]}$ and $R^{[+]}$.  
This choice is driven by the fact that the  $(n+1)$-dimensional  faces are not in a one-to-one relation with the $(n-1)$-dimensional faces. 
Given these considerations we propose the following explosive higher-order Kuramoto dynamics:
\bea
\dot {\bm{\theta}}&=&\bm{\omega}- \sigma  R^{[-]}{\bf B}_{[n+1]}\sin {\bf B}^{\top}_{[n+1]}\bm{\theta}\nonumber \\
&&-\sigma   R^{[+]}{\bf B}^{\top}_{[n]}\sin {\bf B}_{[n]}\bm{\theta},
\eea
valid for each simplex $\alpha\in S_n$. This dynamics  can be projected on the dynamics of $n+1$ and $n-1$ dimensional simplices producing now two  equations coupled by the global order parameters $R^{[+]}$ and $R^{[-]}$:
\bea
{{\dot{\bm{\theta}}^{[+]}}}&=&{\bf B}^{\top}_{[n+1]}\bm{\omega}-\sigma R^{[-]}{\bf L}_{[n+1]}^{[down]}\sin ({\bm{\theta}^{[+]}}),\nonumber \\
{{\dot{\bm{\theta}}^{[-]}}}&=&{\bf B}_{[n]}\bm{\omega}-\sigma R^{[+]} {\bf L}_{[n]}^{[up]}\sin ({\bm{\theta}^{[-]}}).
\eea
We have simulated  the explosive higher-order Kuramoto  dynamics on simplices of dimension $n=1$ on the configuration model of simplicial complexes with power-law distribution of generalized degrees.

A discontinuous phase transition emerges in $R^{[+]}$ and $R^{[-]}$ (see  Fig. $\ref{fig:coupled}$).
This transition is also reflected  on the irrotational and solenoidal components of the dynamics on the $n$-dimensional phases captured by the order parameters $R^{[1]}$ and $R^{[2]}$, while due to the presence of the uncoupled harmonic component $R$ remains close to zero (see  Fig. $\ref{fig:combined}$). The nature of the phase transition does not change significantly if we consider simplicial complexes with Poisson generalized degree distribution (see SM \cite{SI}). Our analytical framework (see SM  \cite{SI}) explains the physics behind this discontinuous phase transition.

{\it Conclusions - }
We have introduced the higher-order Kuramoto dynamics designed to characterize the coupling between phases associated with higher-dimensional simplices, such as links, triangles and so on.
This framework  allows us to define a topologically projected dynamics  on faces of dimension $n-1$ and $n+1$, which obey a dynamics of coupled oscillators. 
We have considered two versions of the higher-order Kuramoto dynamics, the simple and the explosive higher-order Kuramoto dynamics, and we have simulated them on the  simplicial complex configuration model.
We have found that the simple higher-order Kuramoto dynamics
displays  continuous phase transitions for the projected dynamics defined on $n+1$ and $n-1$ faces.
Interestingly, however, when we introduce a coupling between the dynamics projected on the $n+1$ and $n-1$ dynamical phases, as 
in the explosive higher-order Kuramoto dynamics, the system then displays an  explosive synchronization transition.
This work opens innovative perspectives in characterizing the Kuramoto dynamics on higher dimensional simplices, and it shows that a higher-order synchronization dynamics defined on $n$-dimensional simplices (as for example  links) can induce a simultaneous discontinuous transition on its projected dynamics defined  on  $(n-1)$ and $(n+1)$-dimensional simplices (i.e. nodes and triangles).
In the future the proposed  dynamical model can be extended in different directions. For instance one could explore  coupling the dynamics of faces  of different dimensions  or 
other mechanisms leading to explosive synchronization.  

This work is partially supported by SUPERSTRIPES Onlus.
This research utilized Queen Mary's Apocrita HPC facility, supported by QMUL Research-IT. http://doi.org/10.5281/zenodo.438045. G.B. thanks Ruben Sanchez-Garcia for interesting discussions and for sharing his code to evaluate the high-order Laplacian.
A.P.M. and  J.J.T. acknowledge financial support from the Spanish Ministry of Science and Technology, and the Agencia Espa\~nola de Investigaci\'on (AEI) under grant FIS2017-84256-P (FEDER funds).

\renewcommand\theequation{{S-\arabic{equation}}}
\renewcommand\thetable{{S-\Roman{table}}}
\renewcommand\thefigure{{S-\arabic{figure}}}
\setcounter{equation}{0}
\setcounter{figure}{0}
\setcounter{section}{0}

\onecolumngrid
\vspace{0.7cm}
\begin{center}
{\Large\bf             SUPPLEMENTAL MATERIAL}
\end{center}
\section{Additional background material on topology}
\subsection*{Boundary map and incidence matrix}
In topology the $n$-chains  are the elements of a free abelian group ${\mathcal C}_n$ with basis the $n$-dimensional simplices of a simplicial complex. The boundary map is a  linear map $\partial_n:{\mathcal C}_n\to {\mathcal C}_{n-1}$ defined by its action on each simplex $\alpha=[i_0,i_1,i_2\ldots,i_n]$. Specifically, the boundary map  associates  every  $n$-dimensional simplex $\alpha=[i_0,i_1,i_2\ldots,i_n]$ with a linear combination of the $(n-1)$-dimensional oriented faces at its boundary, given by 
\bea
\partial_n [i_0,i_1\ldots,i_n]=\sum_{p=0}^n(-1)^p[i_0,i_1,\dots,i_{p-1},i_{p+1},\dots,i_n].
\label{boundary}
\eea
Considering as a base of the space ${\mathcal C}_n$ an ordered set of the $n$-dimensional simplices $\alpha \in \{1,2,\ldots, N_{[n]}\}$, and as a base the space ${\mathcal C}_{n-1}$ an ordered set of the $(n-1)$-dimensional simplices $\beta \in \{1,2,\ldots, N_{[n-1]}\}$,  we can define the $N_{[n-1]}\times N_{[n]}$ incidence matrix ${\bf B}_{[n]}$ representing  the boundary map in the given bases.
Therefore, if we take  as a base of ${\mathcal C}_{n}$ the ordered set of all the $(n)$-simplices $\{\alpha_1,\alpha_2,\ldots\alpha_s \ldots\}$ and if we take as a base of ${\mathcal C}_{n-1}$ the ordered set of all the $(n-1)$-simplices $\{\beta_1,\beta_2,\ldots\beta_r \ldots\}$ the action of the boundary map over any arbitrary $n$-dimensional simplex $ \alpha_s=[i_0,i_1\ldots,i_n]$ given by Eq. ($\ref{boundary}$) can be equivalently expressed as  
\bea
\partial_n \alpha_s=\sum_{r=1}^{N_{n-1}} [B_{[n]}]_{sr}\beta_r.
\label{B}
\eea
Therefore, Eq. ($\ref{B}$) together with Eq. ($\ref{boundary}$) and a given choice of the bases of the spaces ${\mathcal C}_{n}$ and ${\mathcal C}_{n-1}$ fully determine the incidence matrices ${\bf B}_{[n]}$.

\subsection*{Proof of the Eq. ($5$) of the main text}

Eq. ($5$) of the main text can be rewritten for convenience as
\bea
\partial_{n-1}\partial_n=0.
\label{boundary2}
\eea
This equation nicely reveals that the ``boundary of the boundary is null".
In order to have an intuition of this result let us consider the simplex $[1,2,3]$.
Its boundary is 
\bea
\partial_2 [1,2,3]=[2,3]-[1,3]+[1,2]
\eea
Now by noticing that the boundary of a link $[i,j]$ is 
\bea
\partial_1[i,j]=[j]-[i]
\eea
and substituting this expression in $\partial_1\partial_2[1,2,3]$ we obtain
\bea
\partial_1\partial_2[1,2,3]=\partial_1\left([2,3]-[1,3]+[1,2] \right)=0.
\eea
By using the definition of the boundary map, Eq. (\ref{boundary2}) can be derived in full generality.
Indeed, we have 
\bea
\partial_{n-1} \partial_n [i_0,i_1,\ldots, i_n]&=&\sum_{p<r}(-1)^{p+r}[i_0,i_1,\ldots i_{p-1},i_{p+1}\ldots i_{r-1},i_{r+1}\ldots i_n]\nonumber 
\\&&+\sum_{p>r}(-1)^{p+r-1}[i_0,i_1,\ldots i_{r-1},i_{r+1}\ldots i_{p-1},i_{p+1}\ldots i_n]=0.
\eea

\section{Kuramoto dynamics expressed in terms of the incidence matrix}

The Kuramoto dynamics defined on the nodes $i=1,2,\ldots, N_{[0]}$ of the network  is usually expressed  by the dynamical system of equations for the phases $\theta_i$ as
 \bea
\dot {\theta}_i=\omega_i+\sigma \sum_{j=1}^N a_{ij}\sin \left(\theta_j-\theta_i\right),
\label{KG}
\eea
where $a_{ij}$ is the generic element of the adjacency matrix of the network, $\omega_i$  indicates the internal frequency of the nodes and $\sigma$ indicates the  coupling constant among linked oscillators.
However, this system of equations can be equivalently expressed in terms of the  incidence matrix  ${\bf B}_{[1]}$ as
\bea
\dot {\theta}_i=\omega_i-\sigma \sum_{\ell\in S_{1}} [B_{[1]}]_{i\ell}\sin \left(\sum_{j\in S_{0}}[B_{[1]}^{\top}]_{\ell j}\theta_j\right),
\label{KB}
\eea
where  we indicate with $S_{n}$ the set of all simplices of dimension $n$ (of cardinality $|S_{n}|=N_{[n]}$) present in the simplicial complex under consideration. 
Note that Eq. (8) of the main text is equivalent to Eq. (\ref{KB}) the only difference being that Eq.(8)  is expressed in matrix notation. 
We recall that the matrix element of the incidence matrix ${\bf B}_{[1]}$ can be written as 
\bea
[B_{[1]}]_{i\ell}=\left\{\begin{array}{cc} -1 &\mbox{if} \ \ell=[i,j]\\
1 &\mbox{if} \ \ell=[j,i]\\
0&\mbox{otherwise}\end{array}\right.
\eea
Therefore, if we consider the particular link $\ell=[i,j]$  we can express part of equations Eq. (\ref{KB}) as 
\bea
[B_{[1]}]_{i \ell}\sin \left(\sum_{j\in S_{0}}[B_{[1]}^{\top}]_{\ell j}\theta_j\right)=-a_{ij}\sin(\theta_j-\theta_i).
\eea
Equivalently, if we consider the same  link with opposite orientation $\ell=[j,i]$ we can express the same term in  equations Eq. (\ref{KB}) with the same final expression as 
\bea
[B_{[1]}]_{i \ell}\sin \left(\sum_{j\in S_{0}}[B_{[1]}^{\top}]_{\ell j}\theta_j\right)=a_{ij}\sin(\theta_i-\theta_j)=-a_{ij}\sin(\theta_j-\theta_i).
\eea
Since the incidence matrix ${\bf B}_{[1]}$ has non-zero elements only among nodes and links incident to each other, it follows that  Eq. (\ref{KB}) is equivalent to Eq. (\ref{KG}).

\section{Higher-order Kuramoto dynamics on small simplicial complexes}

\begin{figure}[htb!]
\includegraphics[width=0.95\columnwidth]{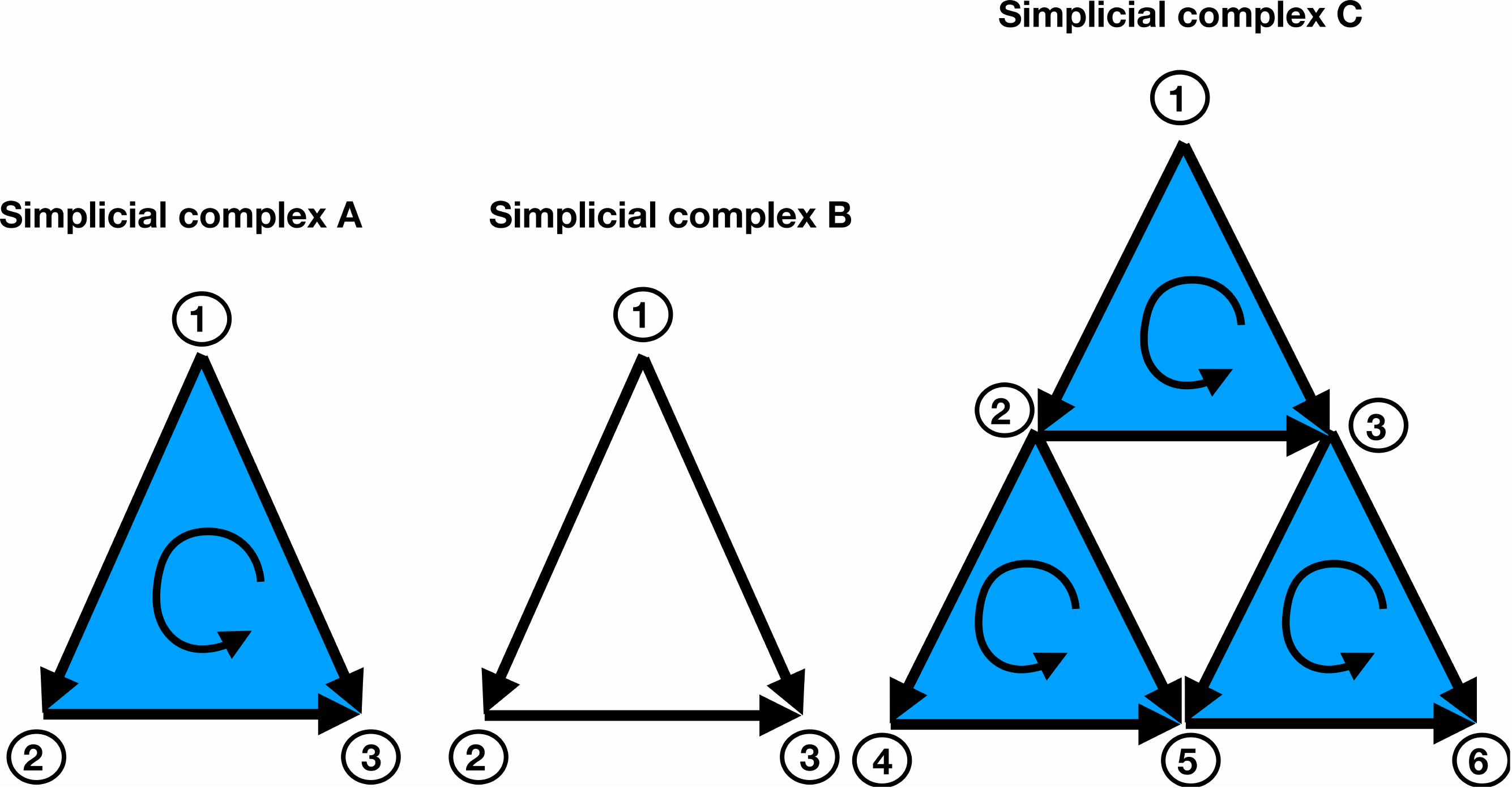}	\caption{Three  small simplicial complexes (simplicial complex A, B and C) considered in this Supplemental Material. 
}
\label{fig:examples}
\end{figure}

In this section we consider the simple higher-order Kuramoto dynamics on small simplicial complexes. In these simplicial complexes it is not possible to see a true thermodynamic phase transition to a synchronized state but their dynamical study can elucidate some fundamental properties of the considered dynamics.
We have considered three simple simplicial complexes of dimension $d=2$ (see Fig. $\ref{fig:examples}$): a full triangle (simplicial complex A), an empty triangle (simplicial complex B), and three incident triangles forming an empty triangle in the middle (simplicial complex C).
The simplicial complex A has trivial homology and Betti numbers $\beta_0=1, \beta_1=0$; the simplicial complexes B and C have instead Betti numbers $\beta_0=\beta_1=1$.
Moreover, the simplicial complex A includes only one $2$-dimensional simplex; the simplicial complex B does not contain any $2$-dimensional simplices, whereas simplicial complex C contains three $2$-dimensional simplices.

The simple higher-order Kuramoto dynamics reads for any  phase $\theta_{\alpha}$ associated to a $n$-dimensional simplicial complex $\alpha$ as
\bea
\dot {\theta}_{\alpha}&=&\omega_{\alpha}-\sigma \sum_{\beta\in S_{n+1}} [B_{[n+1]}]_{\alpha\beta}\sin \left(\sum_{\alpha'\in S_{n}}[B^{\top}_{[n+1]}]_{\beta\alpha'}\theta_{\alpha'}\right)\nonumber \\
&&-\sigma \sum_{\beta\in S_{n-1}} [B^{\top}_{[n]}]_{\alpha\beta}\sin \left(\sum_{\alpha'\in S_{n}}[B_{[n]}]_{\beta \alpha'}\theta_{\alpha'}\right),
\eea
where we draw the intrinsic frequencies $\omega_{\alpha}$  from a normal distribution with mean $\Omega$ and variance $1$, i.e. $\omega\sim{\mathcal N}(\Omega,1)$.
For instance, 
simplicial complex A is formed by the set nodes $\lbrace [1], [2], [3]\rbrace$, the set of links $\{ [1,2],[1,3],[2,3] \}$ and the set of triangles $\{[1,2,3]\}$. The boundary matrices $B_{[1]}$ and $B_{[2]}$ are given by
\bea
B_{[1]} = \begin{pmatrix}
-1 & -1 & 0 \\ 
1 & 0 & -1\\ 
0 & 1 & 1
\end{pmatrix}.
\quad
B_{[2]} = \begin{pmatrix}
1 \\ -1 \\ 1
\end{pmatrix}
\eea
The simple $(n=1)$-order Kuramoto dynamics for simplicial complex A thus reads 
\bea
\dot {\theta}_{[12]}&=&\omega_{[12]}-\sigma \sin(\theta_{[23]}-\theta_{[13]}+\theta_{[12]})-\sigma\left[\sin(\theta_{[12]}-\theta_{[23]})+\sin(\theta_{[13]}+\theta_{[12]})\right],\nonumber\\
\dot {\theta}_{[13]}&=&\omega_{[13]}+\sigma \sin(\theta_{[23]}-\theta_{[13]}+\theta_{[12]})-\sigma\left[\sin(\theta_{[13]}+\theta_{[12]})+\sin(\theta_{[13]}+\theta_{[23]})\right],\nonumber\\
\dot {\theta}_{[23]}&=&\omega_{[23]}-\sigma \sin(\theta_{[23]}-\theta_{[13]}+\theta_{[12]})-\sigma \left[\sin(\theta_{[23]}-\theta_{[12]})+\sin(\theta_{[13]}+\theta_{[23]})\right],
\eea
where $\theta_{[12]},\theta_{[13]}$ and $\theta_{[23]}$ are the phases associated to the links $[1,2],[1,3]$ and $[2,3]$ respectively and where the internal frequency of the links is indicated with $\omega_{[12]}, \omega_{[13]}$ and $\omega_{[23]}$ using an analogous notation.

\begin{figure}[htb!]
 \includegraphics[width=0.75\columnwidth]{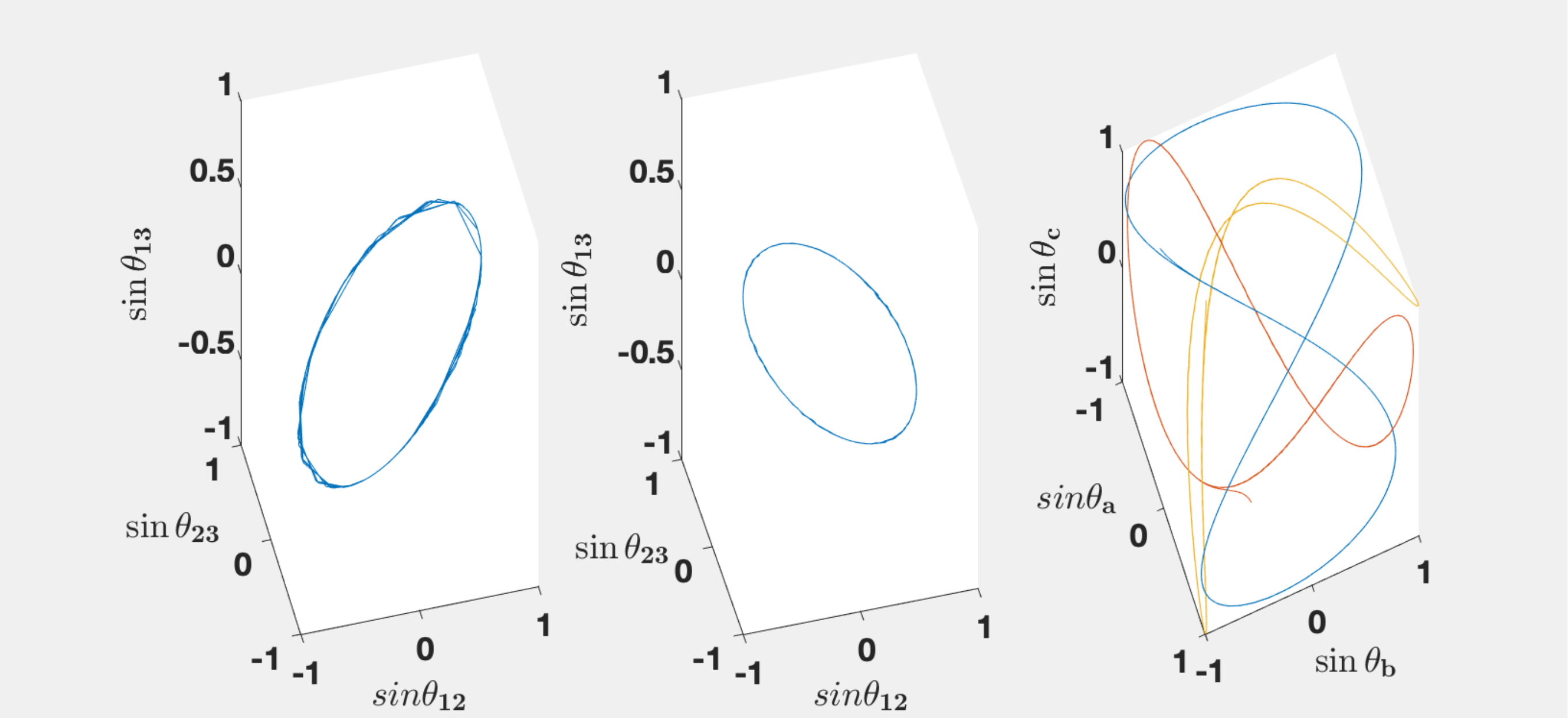}	
 \caption{Examples of dynamics of the simple higher-order Kuramoto model  for the simplicial complexes A (left panel) B (central panel) and C (right panel) with $\Omega=2$. Panel A (and B) show the periodic trajectory of the simple higher-order Kuramoto model on simplicial complex A (B). 
 Panel C shows the trajectory of the simple higher-order Kuramoto model plotting three link variables at the time $\theta_{12},\theta_{13},\theta_{23}$ (blue curve), $\theta_{24},\theta_{25},\theta_{45}$ (red curve), $\theta_{35},\ \theta_{36},\ \theta_{56}$ (yellow curve). 
 The frequencies and initial states used in these plots are indicated in table $\ref{tab:data}$. 
\label{fig:example_dynamics} }
\end{figure}

In figure $\ref{fig:example_dynamics}$ we illustrate the dynamics of the simple higher-order Kuramoto model for these three simpleces, providing evidence that the simple higher-order Kuramoto dynamics on the considered small simplicial complexes can give rise to non-trivial dynamics. In order to do so, we show exemplary trajectories in the phase space defined by the link phases. These examples show the emergence of periodic states under these dynamics.
 However our intention is to provide a more comprehensive analysis of the phase diagram of the simple higher-order  Kuramoto dynamics in a separate publication.

\begin{table}[htb!]
\begin{center}
\begin{tabular}{ |c| c c |c c| }
\hline
& \multicolumn{2}{c|}{SC-A} & \multicolumn{2}{c|}{SC-B}  \\ 
 Simplex  & $\omega_{[ij]}$ & $\theta_{[ij]}$ & $\omega_{[ij]}$ & $\theta_{[ij]}$ \\
 \hline
$[1,2]$ & 2.75 & 5.47 & 1.58 & 2.73 \\  
$[1,3]$ & 1.57 & 6.11 & 1.59 & 2.60 \\
$[2,3]$ & 0.55 & 0.73 & 2.09 & 0.34 \\
\hline
\end{tabular}
\quad
\begin{tabular}{ |c c c || c c c| }
\hline
\multicolumn{6}{|c|}{SC-C}  \\ 
\hline
 Simplex & $\omega_{[ij]}$ & $\theta_{[ij]}$  & Simplex &$\omega_{[ij]}$ & $\theta_{[ij]}$ \\
 \hline
$[1,2]$ & $0.5808$ & $3.8153$ & $[4,5]$ & $0.9793$ & $6.2288$ \\  
$[1,3]$ & $4.5467$ & $6.1627$ & $[3,5]$ & $3.1340$ & $4.7694$ \\
$[2,3]$ & $1.6420$ & $4.5698$ & $[3,6]$ & $2.0385$ & $0.9171$\\
$[2,4]$ & $2.4013$ & $4.5698$ & $[5,6]$ & $2.2309$ & $2.0492$  \\
$[2,5]$ & $1.2938$ & $5.3840$ & - & - & - \\
 \hline
\end{tabular}
\caption{Parameter values used in Fig. $\ref{fig:example_dynamics}$. \label{tab:data}}
\end{center}
\end{table}


\section{The synchronization transition of simple and explosive high-order Kuramoto model}

In this paragraph we provide an analytical framework for the simple and the explosive higher-order Kuramoto models that is able to reveal the main observed phenomenology, albeit being an  approximate approach. In particular the proposed framework reveals the main mechanism that determine the nature of the phase transition (continuous for the simple higher-order Kuramoto model and discontinuous for the explosive higher-order Kuramoto model) and provides an estimate for the critical value of $\sigma=\sigma_c$ in both models. 

\subsection{Simple higher-order Kuramoto model}
We consider the simple higher-order Kuramoto model for the phases $\bm{\theta}$ associated to the $n$-dimensional simplices of the simplicial complex, i.e. 
\bea
\dot {\bm{\theta}}&=&\bm{\omega}-\sigma  {\bf B}_{[n+1]}\sin {\bf B}^{\top}_{[n+1]}\bm{\theta}-\sigma  {\bf B}^{\top}_{[n]}\sin {\bf B}_{[n]}\bm{\theta},
\label{SK}
\eea
where $\bm{\omega}$ is the vector of  intrinsic frequencies $\omega_{\alpha}$  associated to each $n$-dimensional simplex $\alpha$, and  each frequency $\omega_{\alpha}$ is  drawn from a normal distribution with mean $\Omega$ and variance $1$, i.e. $\omega_{\alpha}\sim{\mathcal N}(\Omega,1)$. 
Let us  project the vector of the phases into any harmonic  eigenvector ${\bf u}_h$ of the higher order Laplacian ${\bf L}_{[n]}$ given by 
\bea
{\bf L}_{[n]}={\bf B}^{\top}_{[n]}{\bf B}_{[n]}+{\bf B}_{[n+1]}{\bf B}^{\top}_{[n+1]},
\eea 
obtaining 
\bea
\theta_{h}=\braket{{\bf u}_h|\bm{\theta}},
\eea
where here and in the following we indicate the cross product with the braket notation.
Since the harmonic  eigenvector ${\bf u}_h$ satisfies ${\bf u}_h^{\top}{\bf B}_{[n+1]}={\bf 0}$ and ${\bf u}_h^{\top}{\bf B}_{[n]}^{\top}={\bf 0}$  by using Eq. (\ref{SK}) we get 
\bea
\dot{\theta_{h}}=\braket{{\bf u}_h|{\bf \dot{\theta}}}=\braket{{\bf u}_h|\bm{\omega}}.
\eea
Therefore every harmonic component of the phases $\theta_h$ oscillates with a proper frequency explaning why the order parameter $R$ does not show sign of synchronization, i.e. $R\simeq 0$.
However the component of the phases that are orthogonal to the harmonic component can become synchronized.
This can be observed by considering the variables 
\bea
{\bm{\theta}^{[+]}}&=&{\bf B}^{\top}_{[n+1]}\bm{\theta},\nonumber \\
{\bm{\theta}}^{[-]}&=&{\bf B}_{[n]}\bm{\theta},
\eea
which ``filtrates out" the harmonic component of ${\bm{\theta}}$, i.e.
\bea
\braket{{\bf u}_h|\bm{\theta}^{[+]}}=\braket{{\bf u}_h|\bm{\theta}^{[-]}}=0,
\eea
because by the  theory of the Hodge decomposition it is known that any harmonic eigenvector ${\bf u}_h$  of the higher-order Laplacian ${\bf L}_{[n]}$ is in the kernel of both incidence matrices ${\bf B_{[n+1]}}^{\top}$ and   ${\bf B}_{[n]}$.
 Moreover, ${\bm{\theta}^{[+]}}$ and ${\bm{\theta}^{[-]}}$ are  linearly independent since, according to Hodge decomposition,  the incidence matrices ${\bf B}^{\top}_{[n+1]}$ and ${\bf B}_{[n]}$ are not only simultaneously diagonalizable on the base of eigenvectors $\{{\bf u}_{\lambda}\}$  that diagonalize  the Laplacian ${\bf L}_{[n]}$, but are such that any eigenvector ${\bf u}_{\lambda}$ corresponding to a non zero eigenvalue $\lambda$ of the Laplacian ${\bf L}_{[n]}$ is either in the kernel of   ${\bf B}_{[n+1]}^{\top}$ or in  the kernel of ${\bf B}_{[n]}$, i.e.
 either 
 \bea
{\bf B}_{[n+1]}^{\top}{\bf u}_{\lambda}=0,
 \eea
 or 
 \bea
{{\bf B}_{[n]}\bf u}_{\lambda}=0.
 \eea 
As discussed in detail in the main body of the paper, the simple higher-order Kuramoto model given by Eq.(\ref{SK}) induces the following dynamics on the projected phases ${\bm{\theta}^{[+]}}$ and ${\bm{\theta}^{[-]}}$ 
\bea
{{\dot{\bm{\theta}}^{[+]}}}&=&{\bf B}^{\top}_{[n+1]}\bm{\omega}-\sigma{\bf L}_{[n+1]}^{[down]}\sin ({\bm{\theta}}^{[+]}),\label{tb} \\
{{\dot{\bm{\theta}}^{[-]}}}&=&{\bf B}_{[n]}\bm{\omega}-\sigma{\bf L}_{[n-1]}^{[up]}\sin ({\bm{\theta}}^{[-]}).\label{tb2}
\eea
These equations are independent and have a very similar structure.
Here we focus on the first equation (Eq. $(\ref{tb})$  ) and show that the phases ${\bm{\theta}^{[+]}}$ undergo a continuous synchronization transition with    order parameter  
\bea
R^{[+]}=\left|\frac{1}{N}\sum_{\alpha} e^{i \theta^{[+]}_{\alpha}}\right|.
\label{Rp}
\eea 

Let us consider the base $\{{\bf v}_{\lambda}\}$ of eigenvectors  the down Laplacian ${\bf L}_{[n+1]}^{[down]}$  with  ${\bf v}_{\lambda}$ indicating the eigenvector corresponding to the  $\lambda$  eigenvalue of down Laplacian ${\bf L}_{[n+1]}^{[down]}$.
 By projecting $\bm{\theta}^{[+]}$ on the  basis of the eigenvectors ${\bf v}_{\lambda}$  we obtain 
\bea
\phi^{[+]}_{\lambda}=\braket{{\bf v}_{\lambda}| \bm{\theta}^{[+]}}.
\eea
By using Eq.(\ref{tb}) we can easily see that $\phi^{[+]}_{\lambda}$ obeys the dynamical equation 
\bea
\dot{\phi}_{\lambda}^{[+]}=\braket{{\bf v}_{\lambda}| {\bf B}^{\top}_{[n+1]}\bm{\omega}}-\sigma\lambda \braket{{\bf v}_{\lambda}| \sin ({\bm{\theta}}^{[+]})}.
\eea
The variables ${\phi}_{\lambda}^{[+]}$ reach a stationary state as long as 
\bea
\sigma \lambda \braket{{\bf v}_{\lambda}|\sin ({\bm{\theta}}^{[+]})}=\braket{{\bf v}_{\lambda}| {\bf B}^{\top}_{[n+1]}\bm{\omega}}=\braket{{\bf B}_{[n+1]}{\bf v}_{\lambda}| \bm{\omega}}.
\eea
Since ${\bf v}_{\lambda}$ has the additional property to be   the left eigenvector of ${\bf B}_{[n+1]}$ with eigenvalue $\lambda$ 
we have 
\bea
\braket{{\bf B}_{[n+1]}{\bf v}_{\lambda}| \bm{\omega}}=\lambda \braket{{\bf w}_{\lambda}| \bm{\omega}},
\eea
where ${\bf w}_{\lambda}$ is the right eigenvector of ${\bf B}_{[n+1]}$ corresponding to the eigenvalue $\lambda$.
 Therefore the stationary state for $\dot{\phi}_{\lambda}^{[+]}$ is achieved if 
\bea
\sigma \braket{ {\bf {v}}_{\lambda}|\sin ({\bm{\theta}}^{[+]})}=\braket{{\bf w}_{\lambda}| \bm{\omega}}.
\label{locked}
\eea
as long as $\lambda\neq 0$. 
Since $|\sin (x)|<1$, we obtain that the $\lambda$-eigenmodes with $\lambda\neq 0$ can be locked only if 
\bea
\sigma|\max_{\theta^{[+]}}\braket{{{\bf {v}}_{\lambda}|\sin ({\bm{\theta}}^{[+]})}}|>|\omega_{\lambda}|,
\label{in}
\eea
where $\omega_{\lambda}$ is given by 
\bea
\omega_{\lambda}=\braket{{\bf v}_{\lambda}| \bm{\omega}}.
\label{omega_l}
\eea
From  Eq.(\ref{locked}) it follows that we can express $\sin ({\bm{\theta}}^{[+]})$ as
\bea
\sin ({\bm{\theta}}^{[+]})=\frac{1}{\sigma}\sum_{\lambda \mbox{(locked)},\lambda\neq 0} \omega_{\lambda} {\bf v}_{\lambda} +{\bf y},
\eea
where the first sum is performed only over the locked eigenvalues $\lambda$, 
and ${\bf y}$ is a non specified vector with zero projection on the locked modes, i.e.
\bea
\sum_{\lambda \mbox{(locked)},\lambda\neq0} \braket{{\bf v}_{\lambda}| {\bf y}}=0.
\label{y}
\eea
In order to investigate the synchronization transition on the simple higher-order Kuramoto model   we consider the order parameter $R^{[+]}$ defined in Eq.(\ref{Rp}) and we write it as 
\bea
R^{[+]}=
\left|\frac{1}{N_{[n+1]}}\braket{{\bf 1}|e^{i {\bm{\theta}^{[+}}}}\right|=
\left|\frac{1}{N_{[n+1]}}\sum_{\lambda}\braket{{\bf 1}|{\bf v}_{\lambda}}\braket{{\bf v}_{\lambda}|e^{i {\bm{\theta}^{[+]}}}}\right|,
\eea
where we have used the identity
\bea
1=\sum_{\lambda}\ket{{\bf v}_{\lambda}}\bra{{\bf v}_{\lambda}}.
\eea
Therefore $R^{[+]}$ can be written as 
\bea
R^{[+]}\simeq \left|\frac{1}{N_{[n+1]}}\sum_{\lambda} \braket{{\bf 1}|{\bf v}_{\lambda}} \left[\braket{{\bf v}_{\lambda}|\cos{ {\bm{\theta}^{[+]}}}}+i\braket{{\bf v}_{\lambda}|\sin{ {\bm{\theta}^{[+]}}}}\right]\right|.
\eea
In order to evaluate $R^{[+]}$ we neglect the contribution coming from un-locked eigenmodes, i.e. we neglect ${\bf y}$ in Eq. (\ref{y}) and we get 
\bea
R^{[+]}\simeq \left|\frac{1}{N_{[n+1]}}\sum_{\lambda \mbox{(locked)},\lambda\neq 0} \braket{{\bf 1}|{\bf v}_{\lambda}} \left[\braket{{\bf v}_{\lambda}|\cos{ {\bm{\theta}^{[+]}}}}+i\braket{{\bf v}_{\lambda}|\sin{ {\bm{\theta}^{[+]}}}}\right]\right|.
\eea
We now make the further  assumption that each locked mode contributes a  constant term equal to $a$ to  $R^{[+]}$, therefore we approximate 
\bea
R^{[+]}\simeq aG_{\sigma},
\eea
where $G_{\sigma}$ is equal to the fraction of locked eigenmodes satisfying the equation Eq.(\ref{in}), i.e.
\bea
G_{\sigma}=\frac{1}{N_{[n+1]}}\sum_{\lambda \mbox{(locked)}\lambda\neq 0} 1.
\eea

\begin{figure}[h!]
\includegraphics[width=0.8\columnwidth]{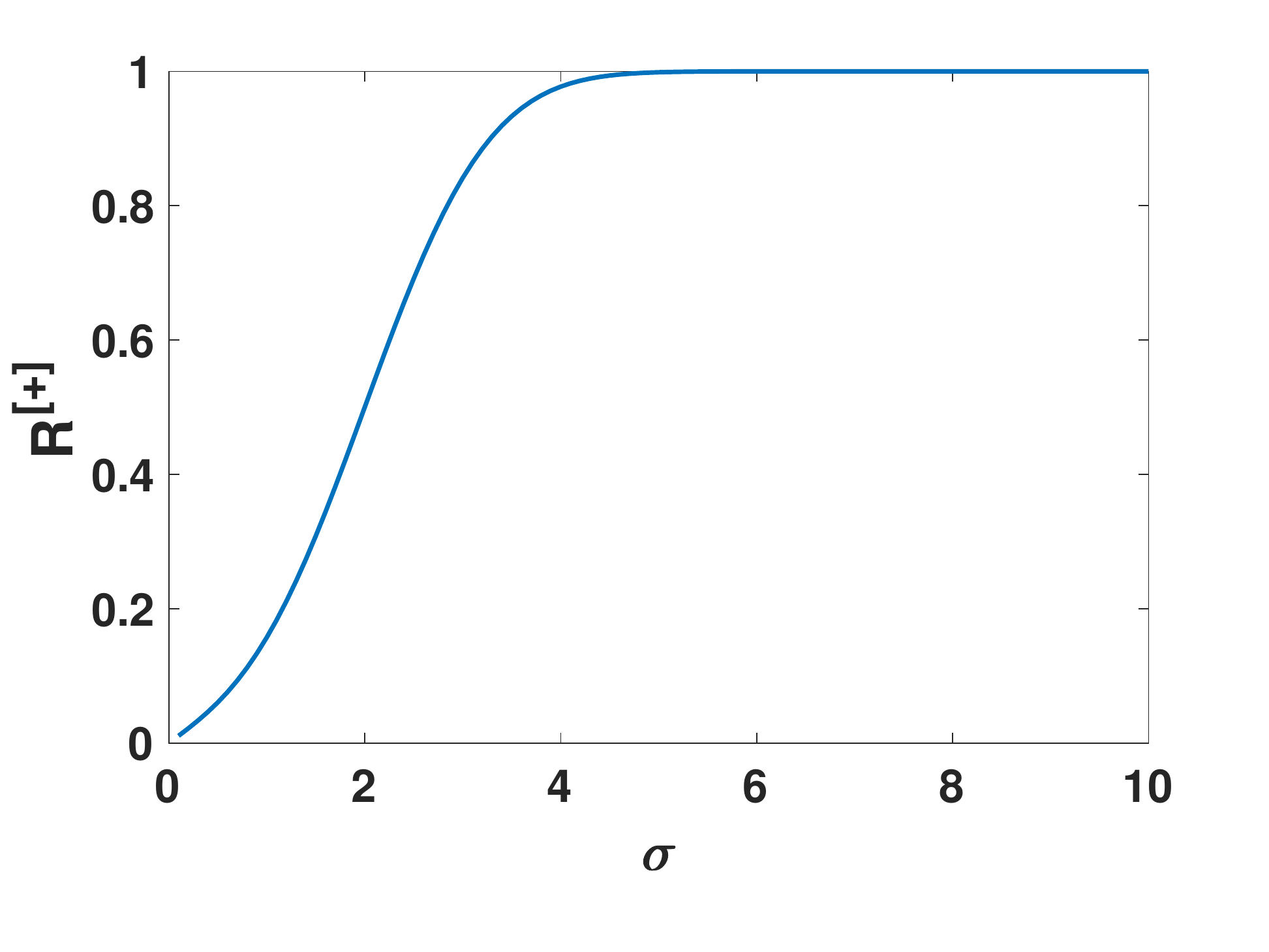}
\caption{Our analytical approach reveals that the simple higher-order Kuramoto model display a continuous phase transition at $\sigma_c=0$. Here we show a typical result for $R^{[+]}$ obtained by plotting Eq. (\ref{scr}) for $\Omega=2$  and $A=B=1$.
}
\label{fig:continuous}
\end{figure}

In order to characterize the transition, we consider Eq. (\ref{in}) determining the condition for an eigenmode to be locked, i.e.
\bea
\sigma\max_{\theta^{[+]}}|{{\bf {v}}_{\lambda}^{\top}\sin ({\bm{\theta}}^{[+]})}|=|{\omega_{\lambda}}|.
\label{in2}
\eea
First of all we notice that $\omega_{\lambda}$ (given by Eq.(\ref{omega_l})), being the projections of Gaussian variables on a orthogonal basis, they are on their turn Gaussian variables with mean $\Omega$ and variance 1, i.e. $\omega_{\lambda}\sim \mathcal{N}(\Omega,1)$.
Secondly we assume that in first approximation
\bea
\max_{\theta^{[+]}}|{{\bf {v}}_{\lambda}^{\top}\sin ({\bm{\theta}}^{[+]})}|\simeq B,
\eea
where $B$ is independent on $\lambda$.
Finally,  by indicating with $b$ the fraction of non-zero eigenvalues of ${\bf B}_{[n+1]}$, we can approximate $G_{\sigma}$ as 
\bea
G_{\sigma}=b\rho_{\sigma},
\eea
where $\rho_{\sigma}$ is the probability that  the $\lambda\neq 0$ eigenmode is locked, i.e.
\bea
\rho_{\sigma}=\frac{1}{\sqrt{2\pi}}\int_{-\sigma B}^{\sigma B} d\omega e^{-(\omega-\Omega)^2/2}=\frac{1}{2}\left[\mbox{erf}\left(\frac{B\sigma-\Omega}{\sqrt{2}}\right)+\mbox{erf}\left(\frac{B\sigma+\Omega}{\sqrt{2}}\right)\right].
\eea
Therefore by putting $A={ab}$ we obtain, within our approximations, 
\bea
R^{[+]}\simeq A\rho_{\sigma}=\frac{A}{2}\left[\mbox{erf}\left(\frac{B\sigma-\Omega}{\sqrt{2}}\right)+\mbox{erf}\left(\frac{B\sigma+\Omega}{\sqrt{2}}\right)\right],
\label{scr}
\eea
which is consistent with a synchronization transition for $R^{[+]}$ obtained at $\sigma_c=0$ for the simple higher-order Kuramoto model,  in agreement with the simulation results.
A similar approach can be used to show that  also $R^{[-]}$  has a transition of $\sigma_c=0$ in the simple higher-order Kuramoto model.
In Figure $\ref{fig:continuous}$ we show a typical curve for the synchronization transition of $R^{[+]}$ obtained using Eq. (\ref{scr}) with $\Omega=2$ and  $A=B=1$.

\subsection{Explosive higher-order Kuramoto model}

For the explosive higher-order Kuramoto model, we start by considering the following equations for the projected phases $\bm{\theta}^{[+]}$ and $\bm{\theta}^{[-]}$
\bea
{{\dot{\bm{\theta}}^{[+]}}}&=&{\bf B}^{\top}_{[n+1]}\bm{\omega}-\sigma R^{[-]}{\bf L}_{[n+1]}^{[down]}\sin ({\bm{\theta}}^{[+]}),\label{tbx} \\
{{\dot{\bm{\theta}}^{[-]}}}&=&{\bf B}_{[n]}\bm{\omega}-\sigma R^{[+]}{\bf L}_{[n-1]}^{[up]}\sin ({\bm{\theta}}^{[-]}).\label{tb2x}
\eea
We now  observe that Eq.(\ref{tbx})  differs from Eq. (\ref{tb}) only by the substitution 
\bea
\sigma\to \sigma R^{[-]},
\eea
and similarly Eq. (\ref{tb2x}) differs from Eq.(\ref{tb2}) only by the substitution 
\bea
\sigma\to \sigma R^{[+]}.
\eea
Therefore, by following the same arguments used in the previous paragraph to derive Eq.(\ref{scr}) we obtain the coupled system of equations for the order parameter of the   $R^{[+]}$ and $R^{[-]}$ of the explosive higher-order Kuramoto dynamics,
\bea
R^{[+]}=F^{[+]}(R^{[-]})=\frac{A^{[+]}}{2}\left[\mbox{erf}\left(\frac{B^{[+]}R^{[-]}\sigma-\Omega}{\sqrt{2}}\right)+\mbox{erf}\left(\frac{B^{[+]}R^{[-]}\sigma+\Omega}{\sqrt{2}}\right)\right], 
\label{unox}\\
R^{[-]}=F^{[+]}(R^{[+]})=\frac{A^{[-]}}{2}\left[\mbox{erf}\left(\frac{B^{[-]}R^{[+]}\sigma-\Omega}{\sqrt{2}}\right)+\mbox{erf}\left(\frac{B^{[-]}R^{[+]}\sigma+\Omega}{\sqrt{2}}\right)\right].
\label{duex}
\eea

\begin{figure}[h!]
\includegraphics[width=0.8\columnwidth]{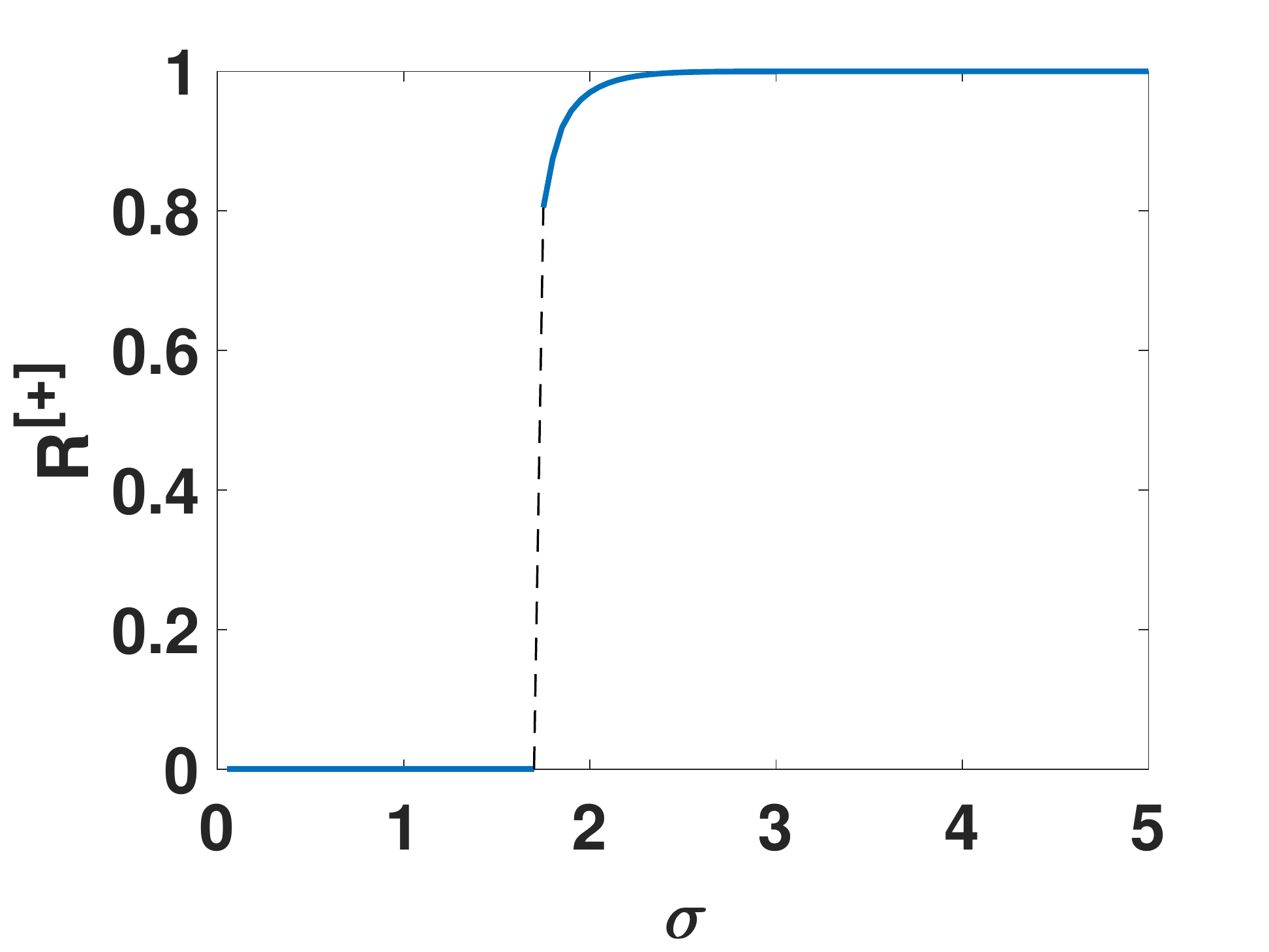}
\caption{ Our analytical approach reveals that for the explosive higher-order Kuramoto model the synchronization transition is discontinuous. Here we show $R^{[+]}$ for a typical scenario obtained by integrating Eq. (\ref{tbx}) and Eq.(\ref{tb2x}) for   $\Omega=2$, $A^{[+]}A^{[-]}=1$ and $B^{[+]}=B^{[-]}=2$. Note that given the symmetric choice of the parameter in this case we will have $R^{[-]}=R^{[+]}$.
}
\label{fig:discontinuous}
\end{figure}

This system of equations display a discontinuous transition. In  Fig. $\ref{fig:discontinuous}$ we show a typical result obtained by integrating numerically Eq. (\ref{tbx}) and Eq.(\ref{tb2x}) for $\Omega=2$, $A^{[+]}=A^{[-]}=1$ and $B^{[+]}=B^{[-]}=2$.
The point of the discontinuous phase transition can be obtained analytically by imposing that the Jacobian matrix of the system of equations 
\bea
R^{[+]}-F^{[+]}(R^{[-]})=0,\nonumber \\
R^{[-]}-F^{[+]}(R^{[+]})=0.
\label{sys}
\eea
(equivalent to the system of Eqs. (\ref{sys})) 
has a determinant equal to zero. In this way we  obtain the equation
\bea
1=\frac{dF^{[+]}(R^{[-]})}{dR^{[-]}}\frac{dF^{[-]}(R^{[+]})}{dR^{[+]}},
\label{J}
\eea
where
\bea
\frac{dF^{[\pm]}(R^{[\pm]})}{dR^{[\mp]}}=\frac{1}{\sqrt{2\pi}}B^{[\pm]}\sigma\left[\exp\left(-\frac{(\Omega-B^{[\pm]}R^{[\mp]})^2}{2}\right)+\exp\left(-\frac{(\Omega+B^{[\pm]}R^{[\mp]})^2}{2}\right)\right].
\eea
The critical point $\sigma_c$ can be  obtained by solving Eqs. (\ref{unox}),(\ref{duex}) together with Eq.(\ref{J}).
For instance, for the case studied in Fig. \ref{fig:discontinuous} we get $\sigma_c=1.7760\ldots$ and $R^{[+]}_c=R^{[-]}_c=0.7981\ldots$.
In general the critical value $\sigma_c$ will be a function of $\Omega$ and of the topology of the simplicial complex, captured  within this approximate solution by the constants $A^{[\pm]}$ and $B^{[\pm]}$.

\section{Higher-order Kuramoto dynamics on large simplicial complexes}

In the main text we  have reported the simulations of  the simple and the explosive higher-order {$(n=1)$} Kuramoto dynamics on the $3$-dimensional simplicial complexes produced by the configuration model with power-law generalized degree distribution of the nodes.
Here we show evidence that the nature of the phase transition does not change if the generalized degree distribution is  more uniform or if the simplicial complex has a  non trivial network geometry.
In particular we consider the simple (Eq. (9) of the main text) and the    explosive (Eq. (12) of the main text) Kuramoto dynamics defined on simplices of order $n=1$ of    a $3$-dimensional simplicial complex produced by the configuration model \cite{Configuration} with Poisson generalized degree distribution of the nodes with average $c=3$. 
These simplicial complexes have Betti numbers $\beta_1>0,\beta_2=0$.
For the simple Kuramoto dynamics we  observe that the projected dynamics on the $2$-dimensional simplices and the $0$-dimensional simplices display a continuous  synchronization transition  whereas for the explosive Kuramoto dynamics we observe a discontinuous transitions (see Figures $\ref{fig:simple}$ and $\ref{fig:combined}$).
  
\begin{figure}[h!]
\includegraphics[width=0.9\columnwidth]{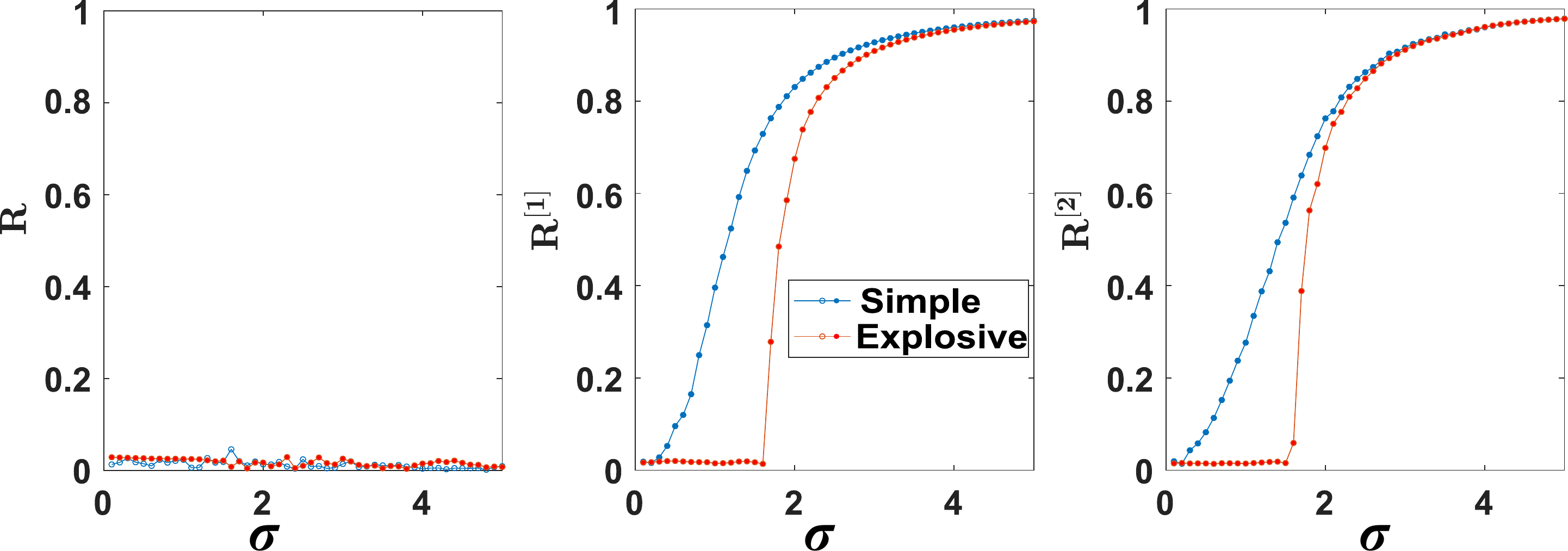}	
\caption{The projection of the  higher-order ($n=1$) Kuramoto dynamics on $(n-1)$-dimensional faces and $(n+1)$-dimensional faces is investigated by plotting the order parameters $R^{[+]}$ {(left panel)} and $R^{[-]}$ (right panel), both for the simple (blue points) and explosive (red points) dynamics.
Here  both the simple and the explosive higher-order Kuramoto model have   $\Omega=2$ and are defined on NGF model with $d=3$, $s=-1$ and $N_{[0]}=1000$ nodes, $N_{[1]}=2994$ links and $N_{[2]}=2992$ triangles. 
}
\label{fig:simpleNGF}\label{fig:coupled}
\end{figure}

\begin{figure}[h!]
\includegraphics[width=0.9\columnwidth]{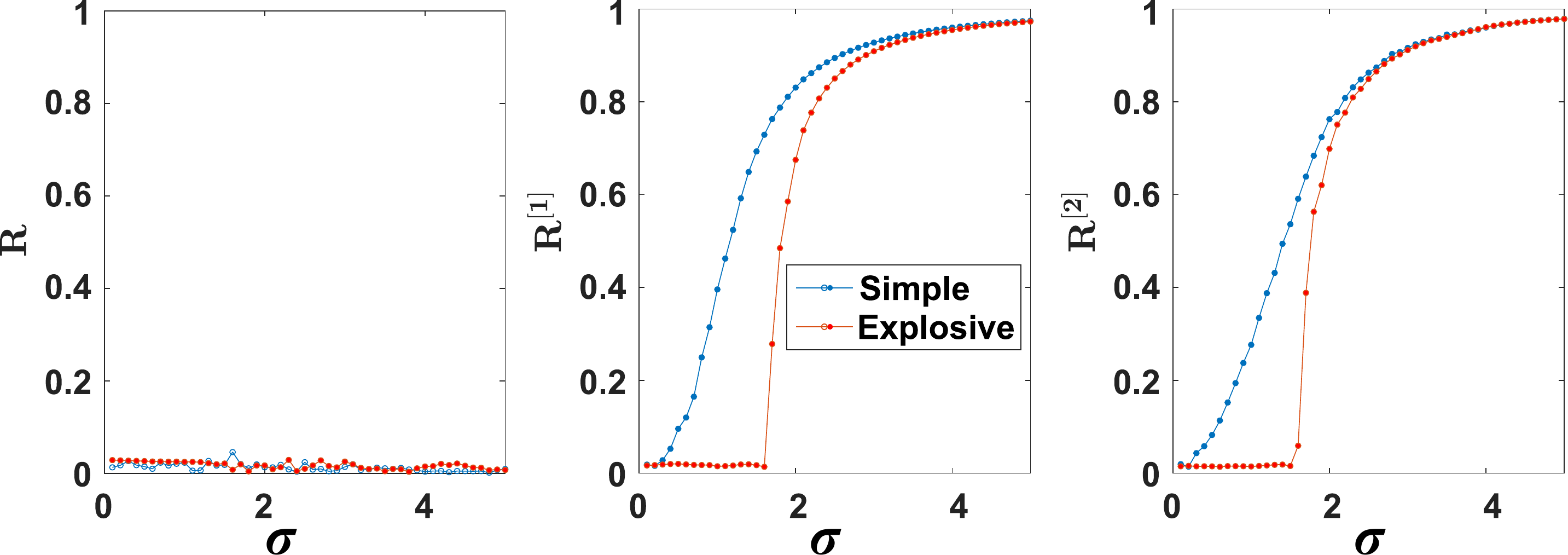}
\caption{The order parameters $R$, $R^{[1]}$ and $R^{[2]}$ of  the simple higher-order ($n=1$)  Kuramoto dynamics {(blue points)} and explosive higher-order ($n=1$)  Kuramoto dynamics {(red points)} are plotted versus the coupling constant $\sigma$.  The network parameters are the same as in Fig. $\ref{fig:simpleNGF}$. 
}
\label{fig:combinedNGF}
\end{figure}

Similar phase diagrams can be observed for the simple Kuramoto model and the explosive Kuramoto dynamics on simplices of order  $n=1$ simulated over a ``Network Geometry with Flavor" \cite{NGF} with $d=3$, $s=-1$ and ``inverse temperature" $\hat{\beta}=0$ (see Figures $\ref{fig:simpleNGF}$ and $\ref{fig:combinedNGF}$).

\begin{figure}[h!]
\includegraphics[width=0.6\columnwidth]{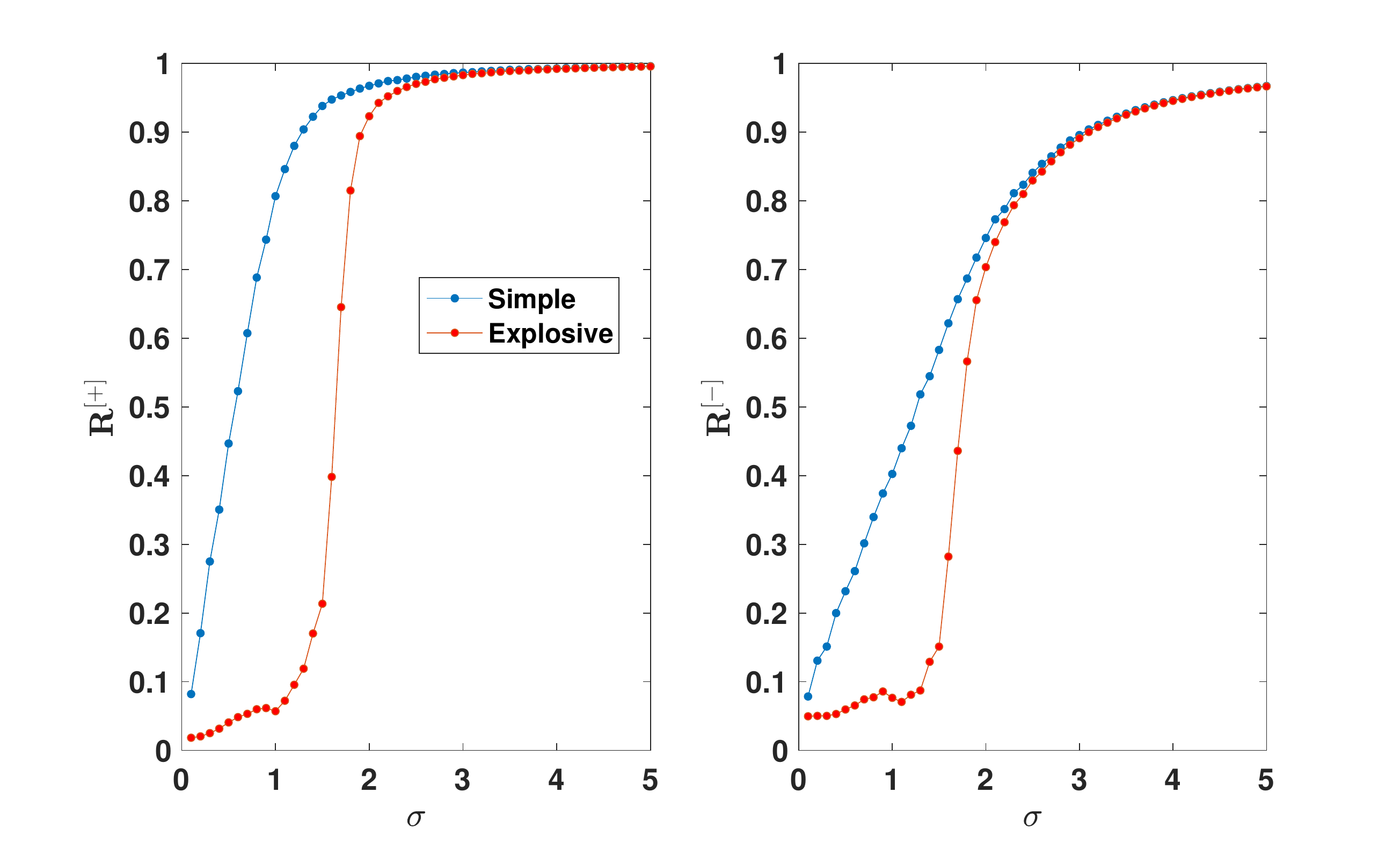}	
\caption{The projection of the  higher-order ($n=1$) Kuramoto dynamics on $(n-1)$-dimensional faces and $(n+1)$-dimensional faces is investigated by plotting the order parameters $R^{[+]}$ {(left panel)} and $R^{[-]}$ (right panel), both for the simple (blue points) and explosive (red points) dynamics.
Here  both the simple and the explosive higher-order Kuramoto model have   $\Omega=2$ and are defined on a configuration model of $N_{[0]}=1000$ nodes, $N_{[1]}=4502$ links and $N_{[2]}=3117$ triangles with generalized degree of the nodes that following a Poisson distribution with average $c=3$. 
\label{fig:simple}\label{fig:coupled}}
\end{figure}

\begin{figure}[h!]
\includegraphics[width=0.9\columnwidth]{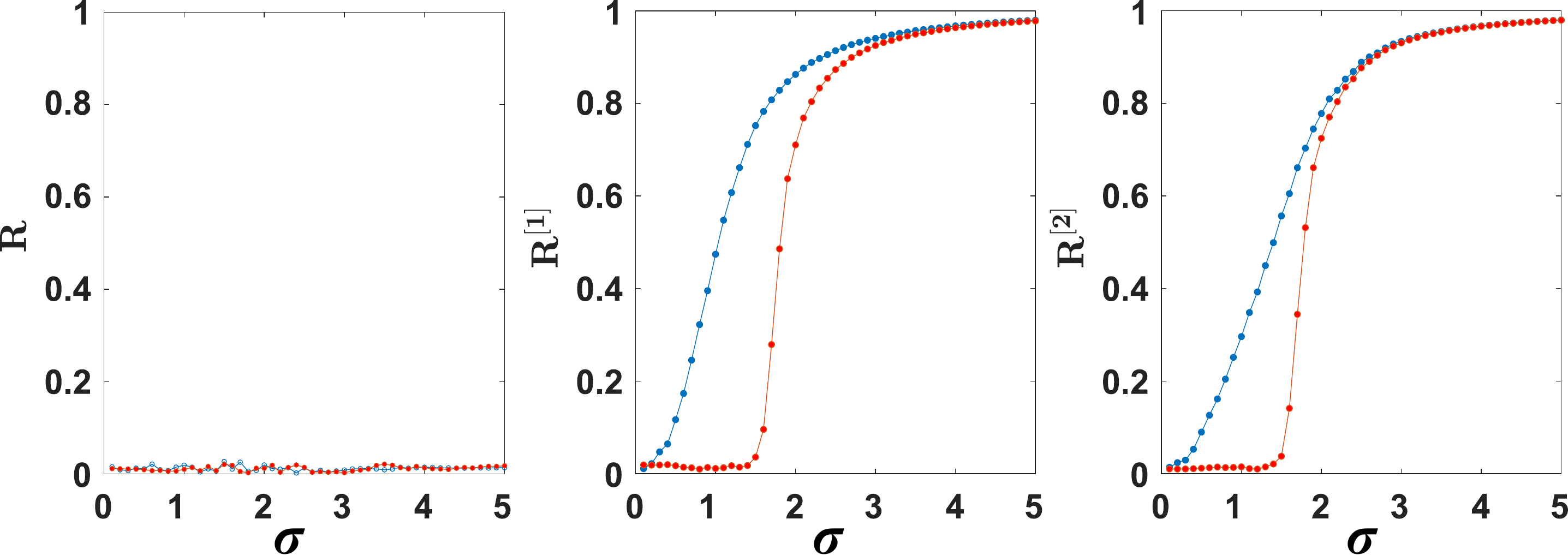}
\caption{The order parameters $R$, $R^{[1]}$ and $R^{[2]}$ of  the simple higher-order ($n=1$)  Kuramoto dynamics {(blue points)} and explosive higher-order ($n=1$)  Kuramoto dynamics {(red points)} are plotted versus the coupling constant $\sigma$.  The network parameters are the same as in Fig. $\ref{fig:simple}$. 
}
\label{fig:combined}
\end{figure}
	
\section{Higher order Kuramoto dynamics on  connectomes of Homo Sapiens and C. elegans}

\begin{figure}[h!]
 \includegraphics[width=0.75\columnwidth]{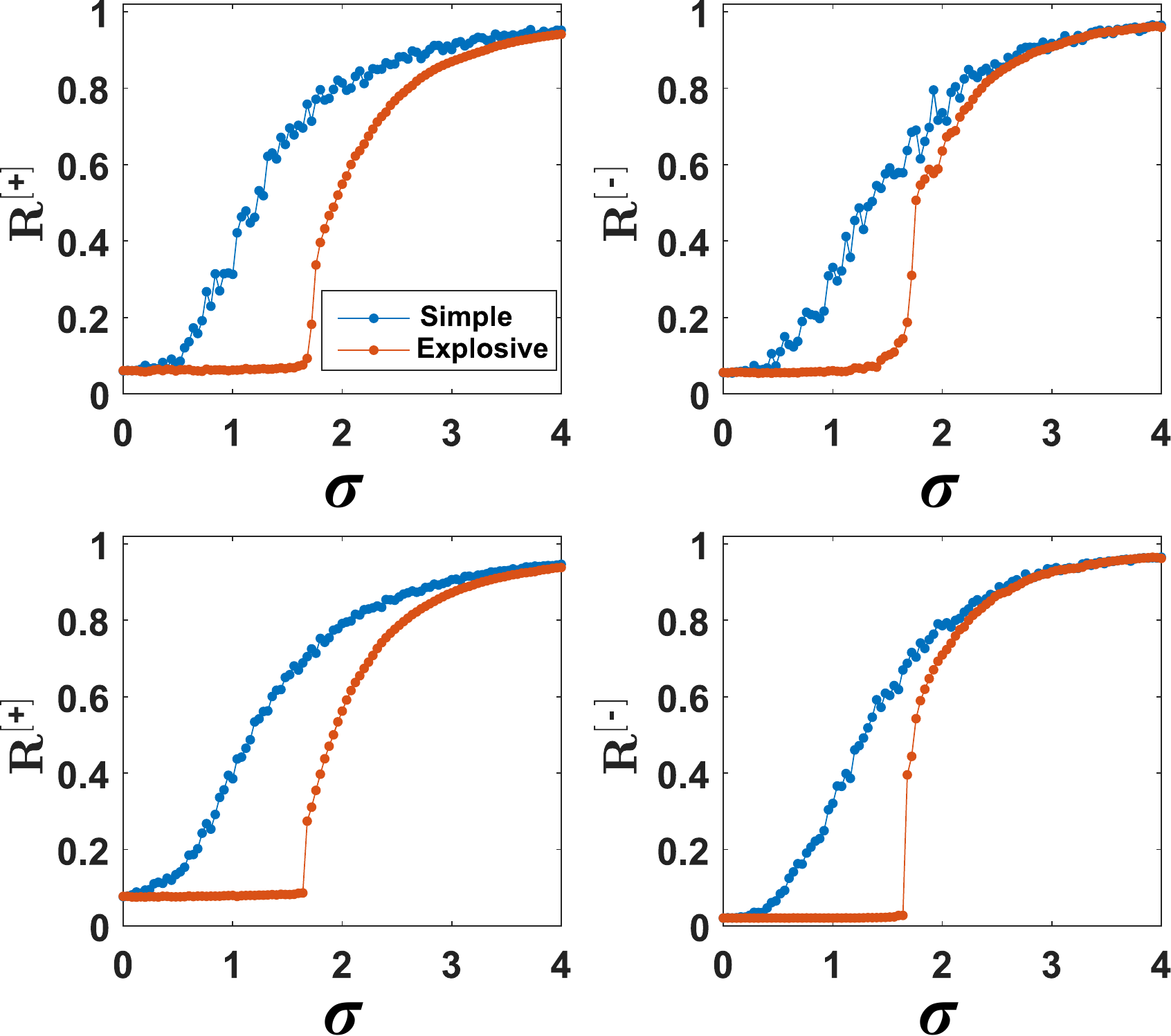}
 \caption{ Here we investigate the higher-order Kuramoto dynamics on top of simplicial complexes extracted from brain connectomes. In particular, we consider the Homo Sapiens connectome  \cite{sapiens} (top panels, $N_{[0]}=66$, $N_{[1]}=254$, $N_{[2]}=291$), and  the C. Elegans connectome  \cite{worm} (bottom panels, $N_{[0]}=277$, $N_{[1]}=1918$, $N_{[2]}=2699$). 
 We investigate the projection of the higher-order ($n=1$) Kuramoto dynamics on $(n-1)$-dimensional faces and $(n+1)$-dimensional faces by plotting the order parameters $R^{[+]}$ {(left panels)} and $R^{[-]}$ (right panels), both for the simple (blue points) and explosive (red points) dynamics. Model parameters are as in previous figures.
\label{fig:connectomes1}}
\end{figure}

\begin{figure}[h!]
 \includegraphics[width=1\columnwidth]{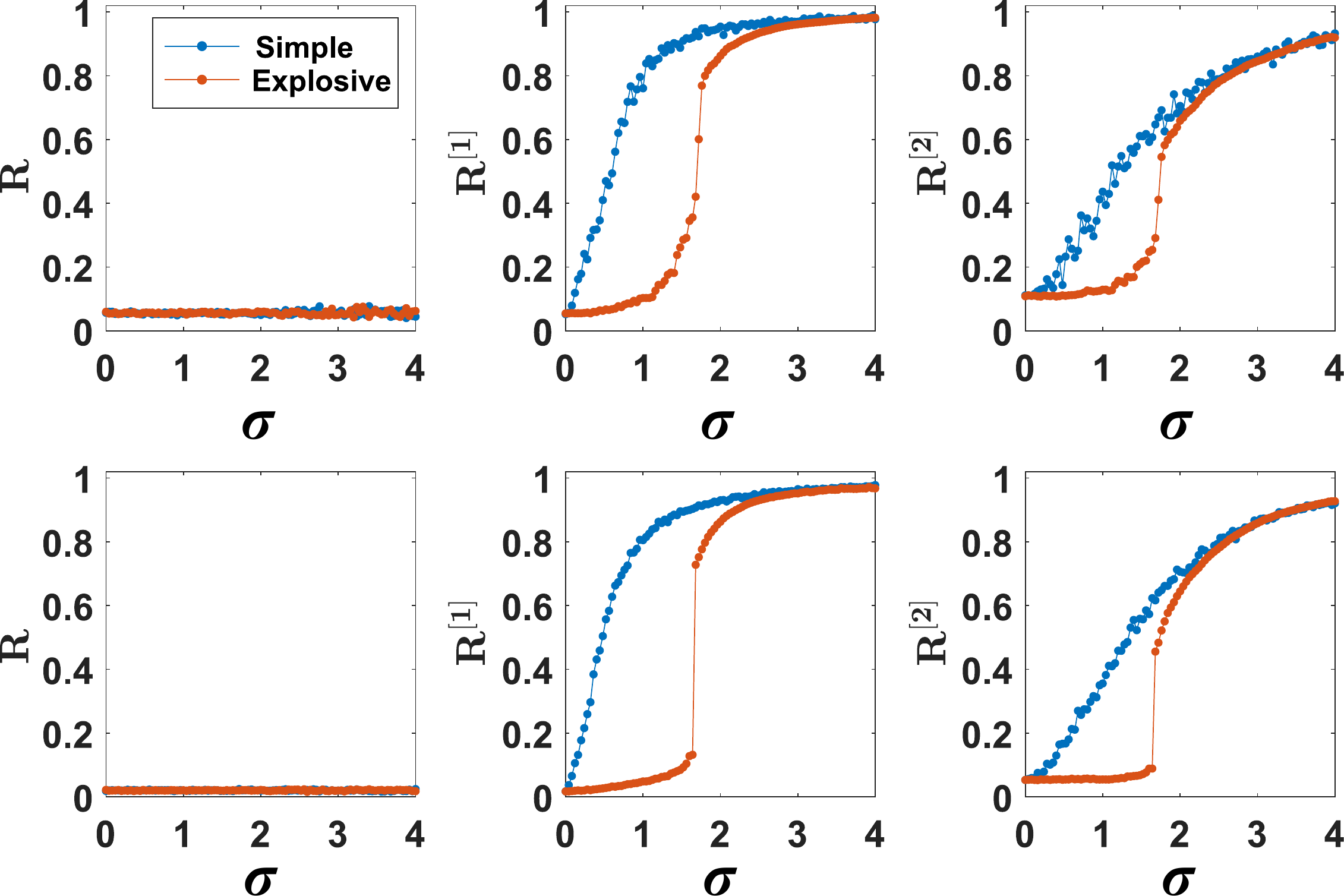}
 \caption{ The order parameters $R$, $R^{[1]}$ and $R^{[2]}$ of  the simple higher-order ($n=1$)  Kuramoto dynamics {(blue points)} and explosive {(red points)} higher-order ($n=1$)  Kuramoto dynamics  are plotted versus the coupling constant $\sigma$, for Homo Sapiens (top panels) and C-Elegans (bottom panels) connectomes. The network parameters are the same as in Fig. $\ref{fig:connectomes1}$. }
\label{fig:connectomes2}
\end{figure}

 In this section we study the synchronization transition of the simple and the explosive higher-order Kuramoto dynamics for $n=1$ taking place over of simplicial complexes constructed starting from real connectomes. In particular,  we consider two connectomes resulting from experimental data and  describing respectively the large-scale connectivity of the Homo Sapiens brain \cite{sapiens} and the microscopic structure of the C. Elegans worm brain \cite{worm}. 
 The brain simplicial complexes  considered in this study are generated starting  from the connectomes an performing the so called clique-complex, i.e. every $(n+1)$ clique is identified with a $n$-dimensional simplex. Therefore every node of the connectome corresponds to a $0$-simplex, every link corresponds to a $1$-simplex and every triangle corresponds to a  a $2$-simplex of the brain simplicial complex.
The simple and the explosive higher-order Kuramoto model have been studied over the brain simplicial complexes and reveal that the nature of the transition of the two dynamical model is unchanged (see Figures $\ref{fig:connectomes1}$ and $\ref{fig:connectomes2}$).  In particular, we observe a continuous phase transition on the dynamics projected on nodes and triangles (see Fig. $\ref{fig:connectomes1}$) for the simple higher-order Kuramoto dynamics and an explosive one for the coupled higher-order Kuramoto dynamics. 
Similarly, in Fig. $\ref{fig:connectomes2}$ we observe that the traditional synchronization parameter $R$ is not appropriate to study the synchronization transition in these higher-order Kuramoto models, whereas the solenoidal and irrotational synchronization parameters $R^{[1]}$ and $R^{[2]}$ display a continuous transition for the simple higher-order Kuramoto dynamics, and an explosive one for the coupled higher-order Kuramoto dynamics.
We also notice that, due to the smaller size of the considered connectomes (and particularly the Homo Sapiens one), the observed discontinuity in the synchronization parameters on the explosive higher-order dynamics at the transition point is smaller than for the simplicial complex models.

Therefore, our results indicate that the nature of the phase transitions is robust and holds even when one consideres simplicial complexes constructed from experimental networks.

\end{document}